\documentclass[12pt]{article}

\usepackage{newtxtext,newtxmath}
\usepackage{tabularx}

\usepackage{graphicx}
\usepackage{booktabs}
\usepackage{subcaption}
\usepackage{makecell}
\usepackage{array}
\usepackage{multirow}
\usepackage[table]{xcolor}
\usepackage{adjustbox}
\usepackage{float}
\usepackage{amsmath}
\usepackage[letterpaper,margin=1in]{geometry}
\usepackage[titletoc]{appendix} 
\usepackage{chngcntr} 

\usepackage[skip=2pt,font=footnotesize]{caption}
\usepackage[percent]{overpic} 

\renewenvironment{abstract}
	{\quotation}
	{\endquotation}

\date{}

\makeatletter
\renewcommand{\fnum@figure}{\textbf{Figure \thefigure}}
\renewcommand{\fnum@table}{\textbf{Table \thetable}}
\makeatother

\usepackage{scicite}

\usepackage{url}

\def\scititle{
	Machine Learning for Electrode Materials: Property Prediction via Composition

}   

\title{\bfseries \boldmath \scititle}

\author{
	Hao Wu$^{1, 2}$, Cameron Hargreaves$^1$, Arpit Mishra$^2$, Gian-Marco Rignanese$^{1,3\ast}$\\
	\small$^{1}$IMCN, Universite catholique de Louvain, 1348 Louvain-La-Neuve, Belgium \and
	\small$^{2}$Avesta Holding, 9400 Ninove, Belgium\and
    \small$^{3}$WEL
    Research Institute, 1300 Wavre, Belgium \and
    \small$^\ast$Corresponding author. Email: gian-marco.rignanese@uclouvain.be \and
}

\begin{document} 

\maketitle

\begin{abstract}
In this work, we benchmark three leading composition-based Machine Learning (ML) frameworks, MODNet, CrabNet, and a random forest model based on Magpie features, predicting the properties of battery electrode materials using the Materials Project Battery Explorer dataset. We evaluate these models based on predictive accuracy, visualize numerical features using two-dimensional embeddings, and quantify performance using standard metrics. Our results demonstrate that CrabNet consistently outperforms the other models across all tests. To validate these findings, we employ bootstrap resampling and two cross-validation (CV) strategies (leave one cluster out and stratified 5-fold CV), comparing each model against a control baseline, using unseen experimental data as a hold-out test. We also apply unsupervised clustering using t-SNE and DBSCAN on physically observed features extracted from matminer, revealing coherent material groupings without prior labels. The final selected model consistently improves over controls, and we believe can be used as a early stage oracle for electrode materials composition screening. Our study aims to identify the error distributions and limitations of the approach, discussing the challenges with developing robust ML models. Despite these constraints, our findings suggest the final selected model is effective for early-stage compositional screening.

\end{abstract}



\section*{Introduction}

The rapid expansion of the portable electronics sector and the global shift toward carbon‑neutral energy storage has directed significant attention towards battery research. The energy density of a battery cell is governed almost entirely by the gravimetric/volumetric capacities of its electrode materials, making the discovery of high‑performance cathodes a top priority for engineers and scientists alike. Consequently, the field has moved from isolated experimental campaigns to an integrated strategy that couples theory, high‑throughput synthesis, and data‑driven discovery \cite{olabi2023rechargeable, sotomayor2019ultra, wu2021building, zhang2018strategy}.

In the 21$^{st}$ century, machine learning (ML) has emerged as the dominant data‑centric paradigm in materials science \cite{wei2019machine}.  With the proliferation of open, high‑quality datasets, such as the Materials Project, OQMD, AFLOWLIB, and the dynamic database of solid-state electrolyte (DDSE), researchers now routinely employ ML to predict a wide range of material properties, from formation energies to elastic moduli \cite{ong2015materials, hargreaves2023database, andersen2021optimade, horton2025accelerated, choudhary2022recent, yang2024dynamic, yang2024user}.  In the context of battery materials, several studies have demonstrated the power of ML: for electrolytes, Wang et al. \cite{wang2025unraveling} presented a data-driven framework relying on large language models for accelerating the discovery of hydride solid-state electrolytes, Wang et al. \cite{wang2025ai} reviewed recent progress in integrating ML, molecular dynamics, and density functional theory within semi-autonomous workflows that accelerate the evaluation of electrolyte materials' properties, Sendek et al. \cite{sendek2018machine} leveraged supervised models to screen for solid-state Li-ion conductors; for electrodes, Zhou et al. \cite{zhou2021machine} used crystal graph neural networks (CGCNN) to identify high-capacity Zn-ion cathodes, and Adam et al. \cite{adam2024navigating} combined data-driven predictions with Bayesian optimization and density functional theory (DFT) to accelerate the design of Li-ion electrodes.

Despite these advances, most electrode‑focused ML workflows still depend on pre‑existing crystal structures, either experimentally measured or DFT‑generated. This requirement limits their applicability to composition‑level screening, which is the natural starting point for high‑throughput exploration.  Ong et al. \cite{ong2015materials} addressed this gap by compiling a comprehensive battery electrode dataset from the Materials Project (MP), yet earlier works such as Zhang et al. \cite{zhang2024interpretable} relied on older and smaller versions of the same data and focused on deep‑learning models that require structural descriptors extracted with tools like matminer \cite{ward2018matminer}.

The efficacy of composition‑level workflows is fundamentally determined by dataset quality. As established in recent studies regarding AI-driven materials discovery and database construction \cite{yang2024dynamic, yang2024user}, the availability of reliable and physically meaningful data is a major bottleneck for ML-based battery prediction. This is due to the simple fact that whilst experimentally validated results are the gold standard, synthesis and characterization is a difficult time-consuming process, we need tremendous quantities of data to compute statistically reliable decision boundaries, and the presence of inaccurate data can push models into making poor predictions, especially in the sparse data regimes we typically work with in materials science. By comparison, high‑throughput databases by nature inherit the systematic approximations of the underlying DFT calculations. Consequently, ensuring the integrity and consistency of this foundational data is paramount, it determines whether a model captures a genuine chemical trend or is following a sequence of computational artefacts.

A rigorous assessment of ML performance on composition‑based electrode prediction is therefore essential. It enables researchers to quantify model uncertainty, benchmark new algorithms against established baselines, and ultimately guide experimental synthesis.  In this study, we evaluate a suite of contemporary and state‑of‑the‑art ML models, including the transformer-based CrabNet, the descriptor-based MODNet architectures, and the magpie descriptor based random forest model, on the MP battery electrode dataset.  We use two standard evaluation metrics to rank the regression models, the mean absolute error (MAE) and the Pearson's $r$-score, as well as using testing errors to obtain a kernel density estimated distribution for the mean and standard deviation model error, allowing us to quantify a models' predictive accuracy and reliability. In addition to using the DFT-computed chemical properties, we have collated a small experimental hold-out dataset to test the final model against. Our results show that while deep learning models excel in capturing complex composition-property relationships, tree‑based ensembles offer comparable performance with greater interpretability and computational efficiency. This benchmark provides a reference point for future work in composition‑level electrode discovery and provide the model weights to integrate our ML approach to other screening regimes. 

\section*{Dataset and Methods}

\subsection*{Dataset}
\begin{figure}[!h]
    \centering

    \begin{subfigure}[t]{0.7\textwidth}
        \begin{overpic}[width=\textwidth]{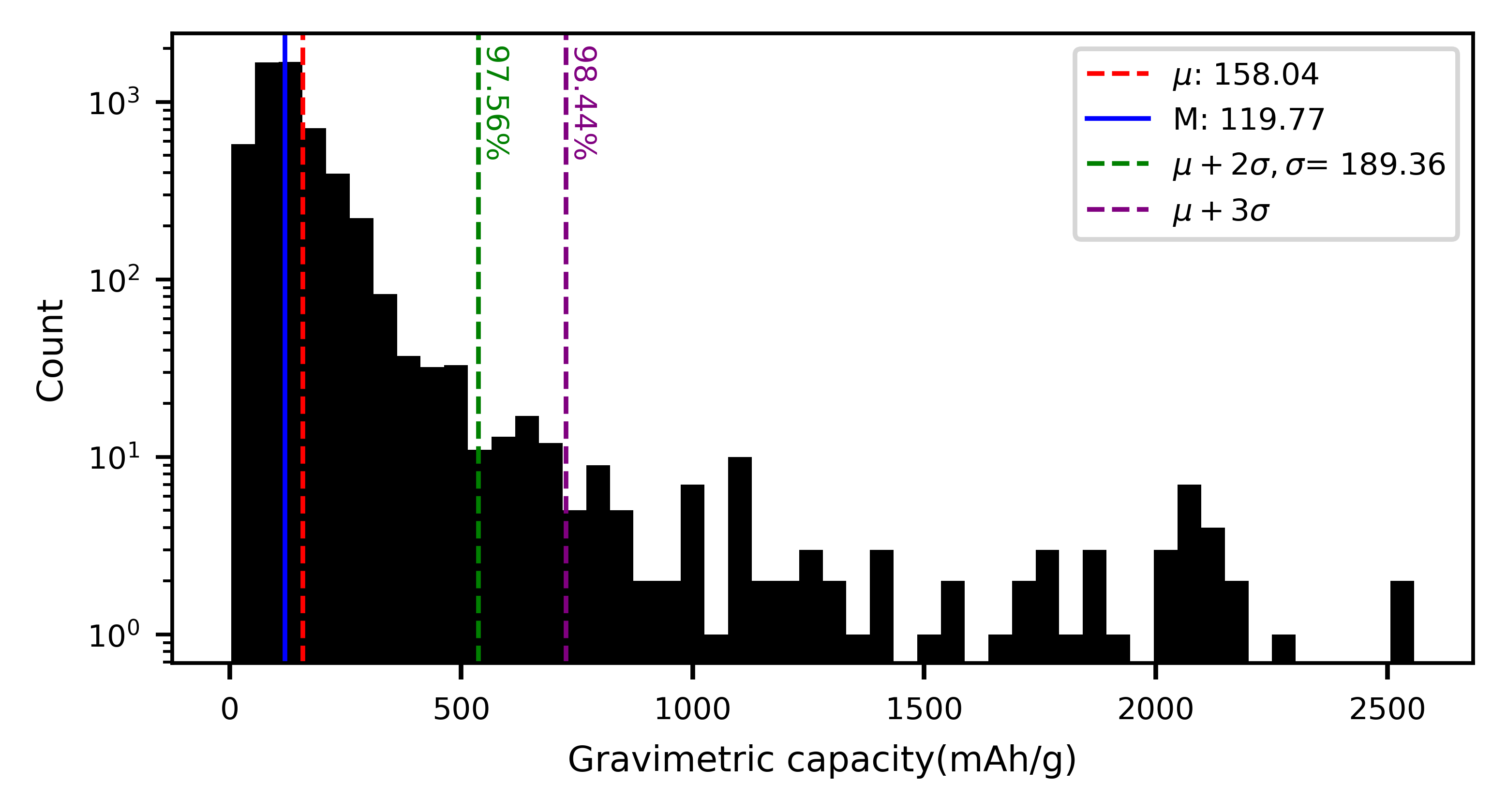}
            \put(2,54){\scriptsize\textbf{(a)}}
        \end{overpic}
    \end{subfigure}

    \vspace{0.2em} 

    \begin{subfigure}[t]{0.7\textwidth}
        \begin{overpic}[width=\textwidth]{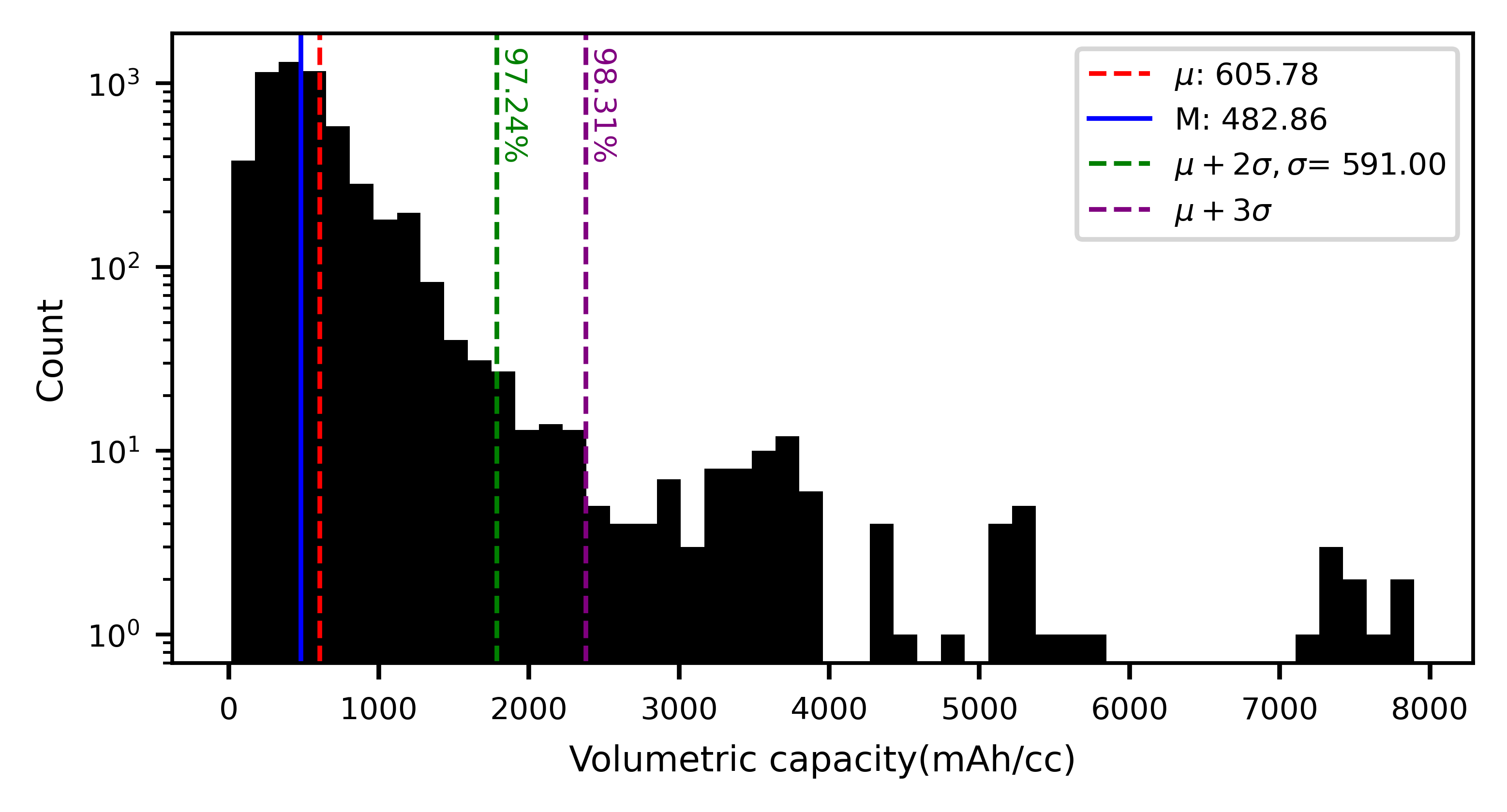}
            \put(2,54){\scriptsize\textbf{(b)}}
        \end{overpic}
    \end{subfigure}

    \vspace{0.2em}

    \begin{subfigure}[t]{0.7\textwidth}
        \begin{overpic}[width=\textwidth]{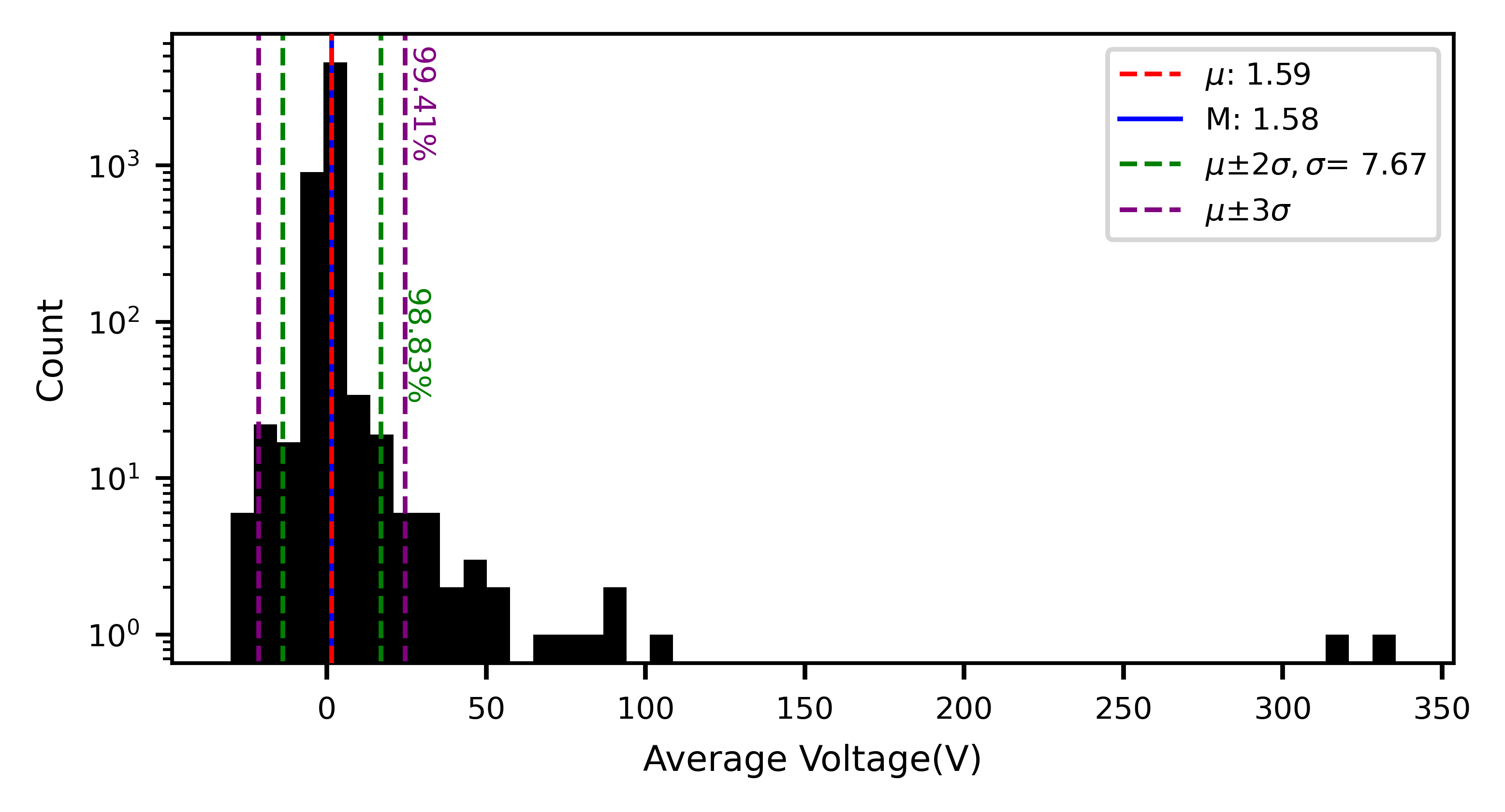}
            \put(2,54){\scriptsize\textbf{(c)}}
        \end{overpic}
    \end{subfigure}

    \caption{Distributions of the target properties in the dataset. Solid blue  and dashed red lines indicate the median (M) and mean ($\mu$) values. Dashed green and purple lines denote empirically observed $\mu$ ± $2\sigma$ (inner green band) and $\mu$ ± $3\sigma$ (outer purple band), with the percentage of the dataset that falls within each interval overlaid. Logarithmic scaling is applied to the count on the y-axis.}
    \label{overlook_value}
\end{figure}

This work uses the Materials Projects Battery Explorer dataset\cite{MPBattery}, a publicly available electrode material dataset from Materials Project (MP)\cite{doi:10.1038/sdata.2015.9}. This dataset has been collated and updated by various research groups \cite{Zhou2004a, Wang2007, Ong2010a}, providing the components and electrochemical characteristics of the electrode materials based on first-principles calculations\cite{ong2013python}. For each electrode material, the dataset provides the MP-ID, the charged and discharged composition, the working ion, and the gravimetric and volumetric capacities, for 5574 electrode materials in total. The distributions of the three properties investigated in this study are shown in Figure  \ref{overlook_value}. 

The calculation of gravimetric and volumetric capacities is based on the following formulae: 
$$D_g = n \cdot e/W_u$$ 
$$D_v = n \cdot e /V_u$$

\noindent where $D_g$ and $D_v$ are the gravimetric and volumetric capacities, $n$ is the number of electrons transferred from the working ion per unit cell, $e$ is the elementary charge, and $W_u$ and $V_u$ are the mass and volume of a single unit cell. 

\begin{figure}[h!]
\centering
\includegraphics[width=0.7\linewidth]{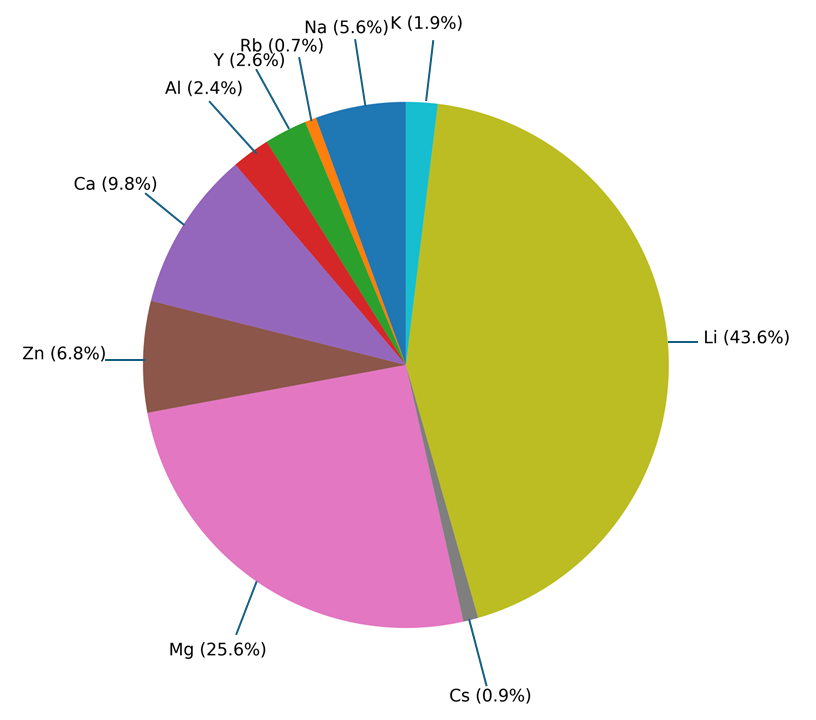}
\caption{Distribution of working ions in the electrode materials dataset.}
\label{overlook_ion}
\end{figure}

Figure \ref{overlook_ion} illustrates the distribution of working ions in the dataset, which reflects contemporary research interests in battery research. Lithium of course dominates, accounting for 43.6\% of the entries, due to its widespread use in rechargeable batteries. Magnesium follows as the second most common working ion at 25.6\%, as there is growing interest in other working ion systems. Whilst this distribution highlights lithium's central role in battery technologies, it also showcases recent efforts to explore multivalent chemistries such as Mg, Ca, Zn, and Al, alongside alternative alkali-ion system like Na and K.

In this work, the gravimetric capacity, volumetric capacity, and average voltage are adopted as regression targets for the ML optimization scheme. A representation of the materials composition at discharge serves as the ML models input, because every element participating in the electrochemical charge–discharge cycle is present at that stage, whereas the charged cathode compositions may not include the working-ion species. As the ML models only take numeric inputs, each of the materials discharged composition must first be featurized. 

\subsection*{Methods}

To featurize each composition, we convert the lexicographic string representing the ratio of elements in the compound into a numeric fixed-length vector. Two main types of featurizing tools have been developed in the literature: chemically-derived and machine-learned ones. The former uses previously reported chemical properties for each constituent element, together with aggregate pooling functions to construct a composition based feature vector. Examples of these include magpie\cite{ward2016general} and JARVIS\cite{choudhary2020joint}; Machine-learned features, such as mat2vec\cite{tshitoyan2019unsupervised} and ElemNet\cite{jha2018elemnet}, are based on the outputs of ML models, typically neural-network architectures, trained to predict known materials properties based on their chemical compositions. When such a model performs well, the high-dimensional, internal representations of the elements can be taken as embeddings for future downstream tasks. 

Once a consistent set of numerical features has been defined to represent the materials compositions as fixed length-vectors, the selected models enter a training phase during which their parameters are optimized to map these input features onto the corresponding target labels. In a subsequent testing phase, these trained models predict the target properties of the test set (unseen data), and the quality of the model is evaluated by quantifying prediction errors. The present study employs three ML models: MODNet \cite{de2021materials}, CrabNet\cite{Wang2021crabnet, Wang2022explainablegap}, and a Random Forest (RF) \cite{heath1993k} using magpie features \cite{ward2016general} (RF@Magpie). 

In MODNet \cite{de2021materials, de2021robust}, the composition-featurization stage concatenates a variety of descriptors drawn from matminer, an extensive repository of physical and chemical features \cite{ward2018matminer}. A feature selection process is then carried forward based on the relationship between raw features and labels, by first calculating the Normalized Mutual Information (NMI)\cite{kraskov2004estimating} between the provided features and the labels, and selecting those with the best relevance-redundancy scores. MODNet then uses a feed-forward neural network to predict the element properties based on the selected features, using a genetic algorithm hyperparameter optimizer. CrabNet \cite{Wang2021crabnet, Wang2022explainablegap} uses a transformer inspired architecture by applying a fractional encoding to mat2vec elemental vector embeddings for the featurizing step to provide a set of numerical vectors. Then, the vectors are processed via a sequence of attention layers and feed-forward layers\cite{vaswani2017attention} to learn and predict the materials' properties. Random Forests (RF) \cite{liu2012new} are a type of ML model that ensembles multiple decision trees to predict materials properties, where the results are generated from the average prediction from each decision tree. In this paper, RF@Magpie combines RF with magpie descriptors. Magpie descriptors \cite{ward2016general} offer a computationally efficient material featurizing approach using a chemically diverse list of attributes, gathered from the literature. 

In this study, the matminer, mat2vec, and magpie embeddings comprise $273$, $199$, and $21$, features respectively. As such high-dimensional spaces are beyond human visualization, dimensionality reduction techniques are applied to obtain two-dimensional (2D) embeddings that reveal the overall distribution of the materials and can also serve as inputs for automated clustering techniques. Three dimensionality reduction methods are employed to visualize the features in 2D: Principal Component Analysis (PCA), t-distributed Stochastic Neighbor Embedding (t-SNE), and Uniform Manifold Approximation and Projection (UMAP). 

PCA is a linear dimensionality-reduction technique that projects high-dimensional features onto a low-dimensional space while maximally preserving the inter-point distances. By aligning the axes along the regions with greatest variance, it can expose dominant patterns in the data, but it can compress high dimensional relationships beyond recognition. Additional PCA results are provided in Appendix Figure \ref{over_ion_dropped}(a, b, c), but this study finds no discernible clustering with projected points falling in overly dense regions, which fail to separate cleanly using automated methods. Moreover, the explained-variance ratios for the first two principal components are below 30\%, indicating that a simple linear projection onto a 2D subspace does not capture the intrinsic structure of these high-dimensional datasets. 

\begin{figure}[h!]
    \centering
    \begin{subfigure}[t]{0.32\textwidth}
        \begin{overpic}[width=\textwidth]{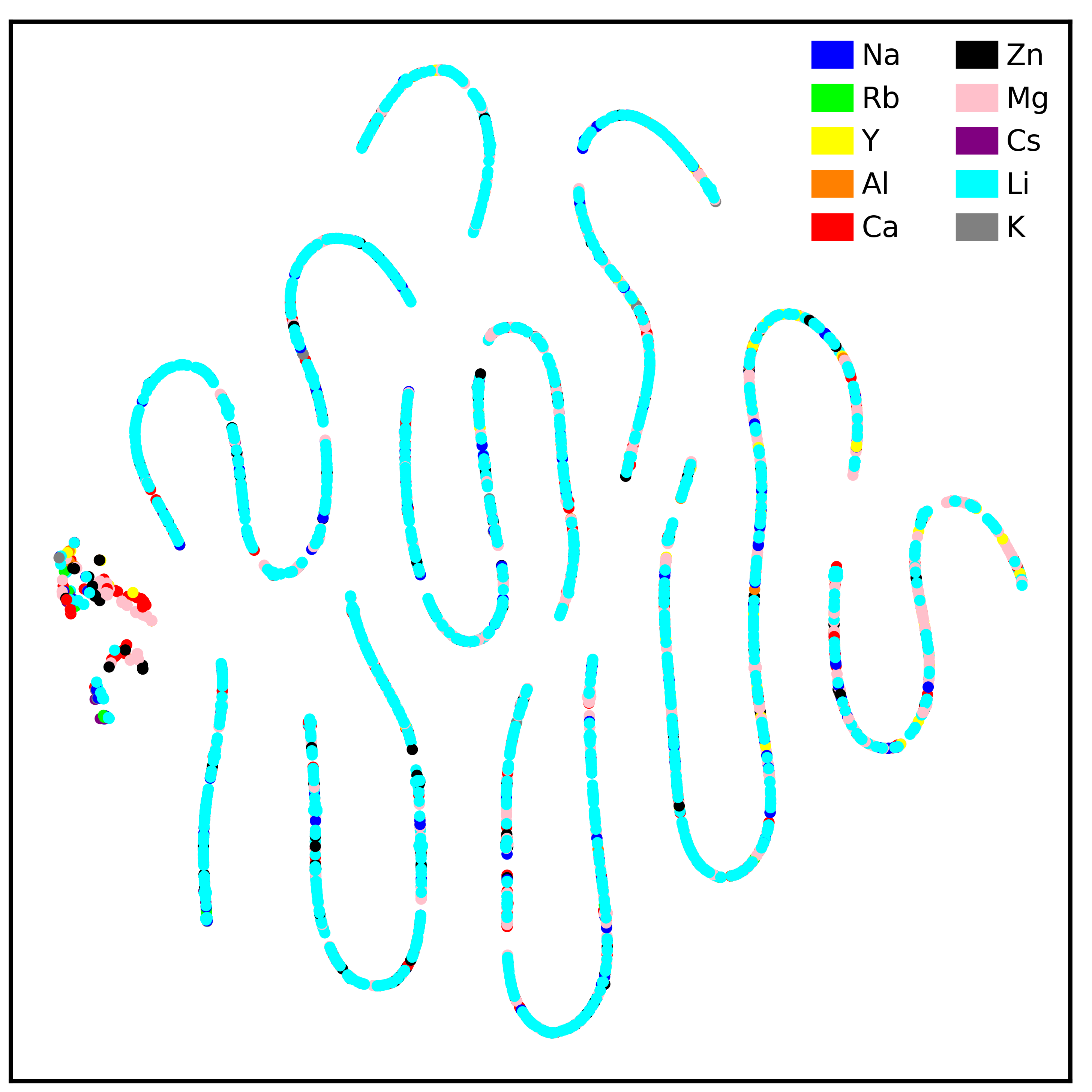}
            \put(0,100){\scriptsize\textbf{(a)}} 
        \end{overpic}
        \label{tsne_modnet_ion}
    \end{subfigure}
    \hspace{-0.3em}
    \begin{subfigure}[t]{0.32\textwidth}
        \begin{overpic}[width=\textwidth]{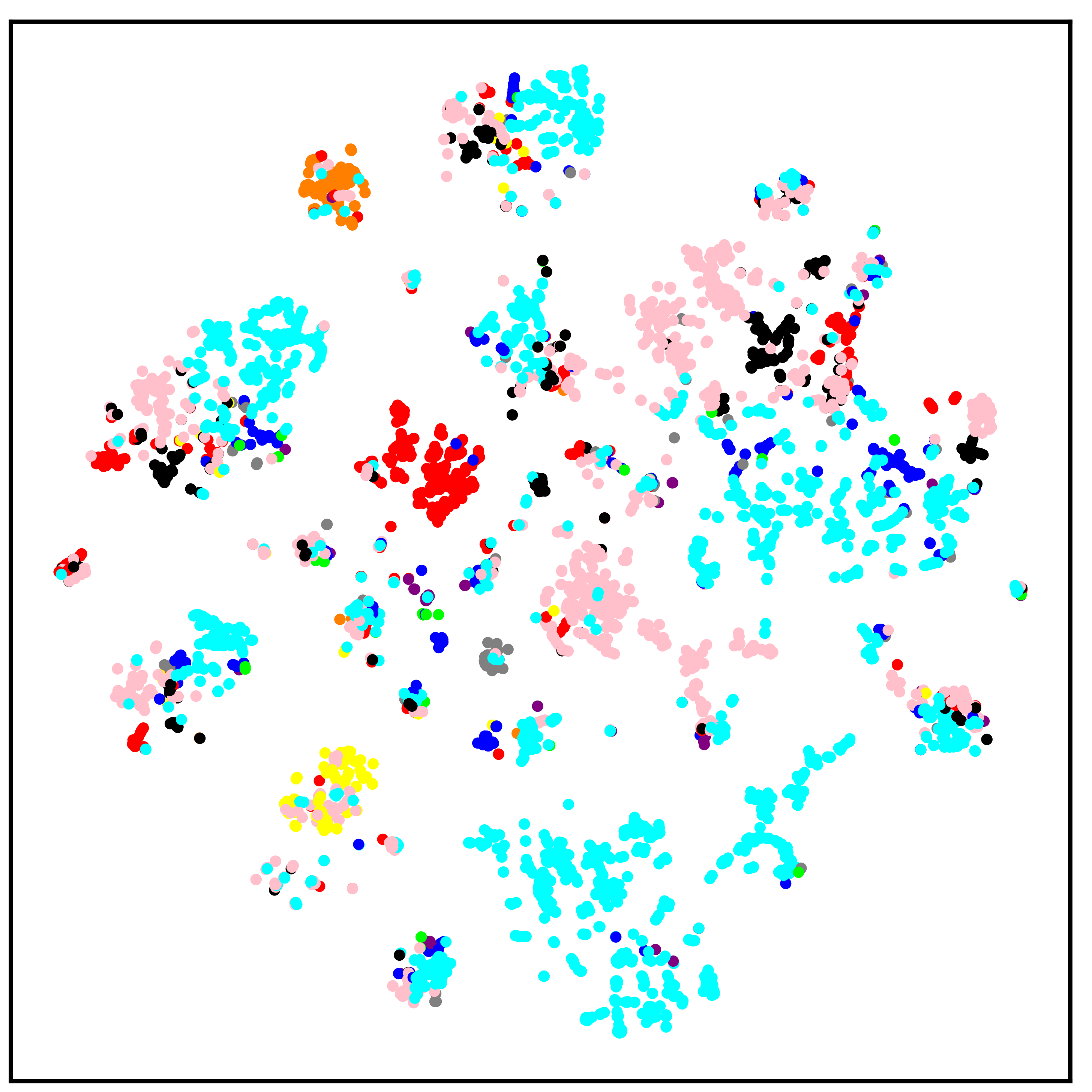}
            \put(0,100){\scriptsize\textbf{(b)}}
        \end{overpic}
        \label{tsne_mat2vec_ion}
    \end{subfigure}
    \hspace{-0.3em}
    \begin{subfigure}[t]{0.32\textwidth}
        \begin{overpic}[width=\textwidth]{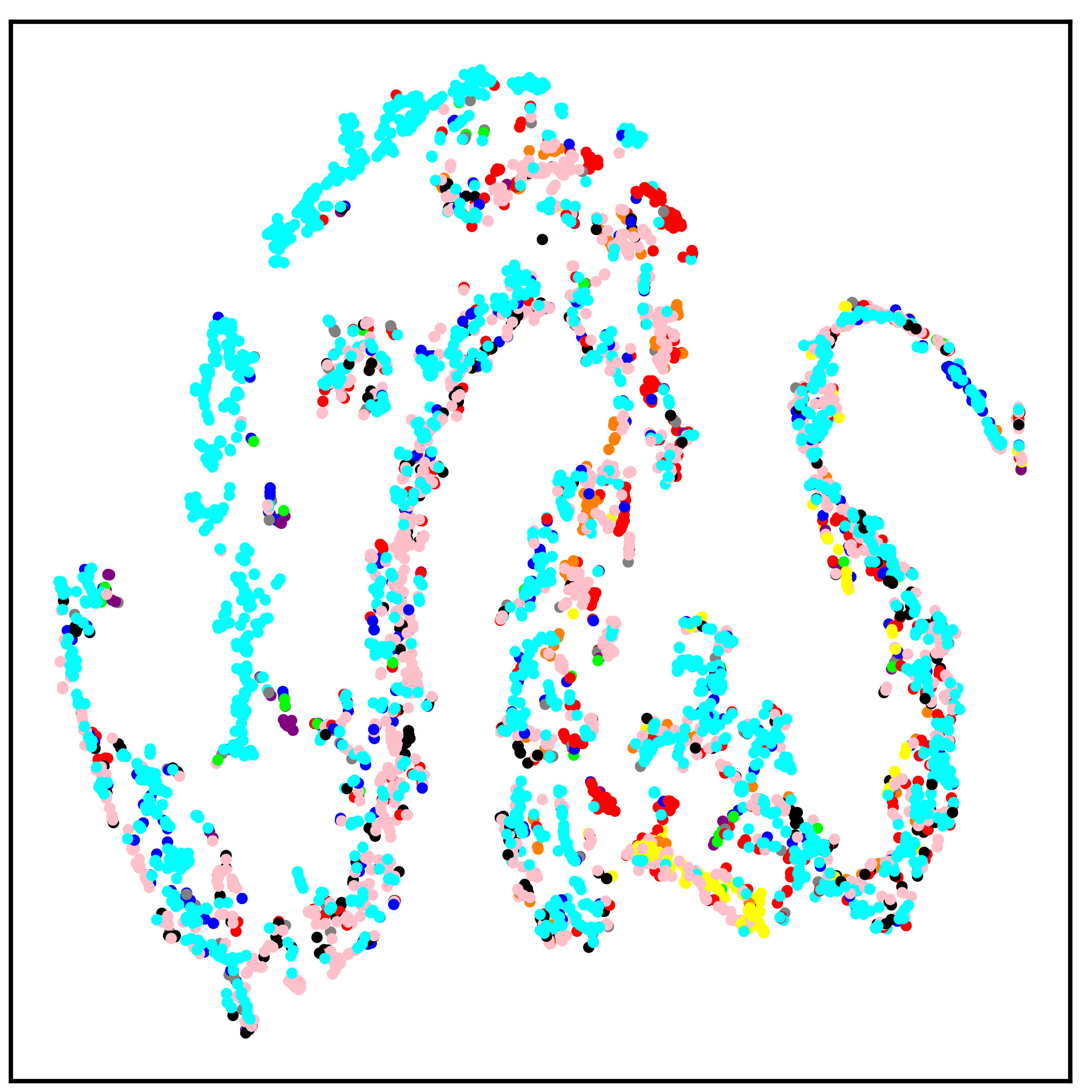}
            \put(0,100){\scriptsize\textbf{(c)}}
        \end{overpic}
        \label{tsne_magpie_ion}
    \end{subfigure}
    \caption{Three t-SNE embeddings of the materials in the dataset, using input features from (a) matminer (used by MODNet), (b) mat2vec (used by CrabNet) and (c) magpie. The points have been colored according to the working ion in the cathode.}
    \label{over_feature}
\end{figure}

Conversely, both t-SNE and UMAP are non-linear dimensionality-reduction methods that focus on preserving local similarity rather than global fidelity. By optimizing local neighborhood pointwise distances, they produce lower‑dimensional embeddings that highlight the local structure even though the global geometry may be more distorted. In this study, both t-SNE and UMAP yield well-separated clusters that align with known chemical relationships. The three UMAP embeddings are provided in Figure \ref{over_ion_dropped}(d, e, f), but as t-SNE uses a similar non-linear approach, and the t-SNE embeddings provide the cleanest separation of meaningful chemical clusters, we focus the discussion on these, which are shown in Figure \ref{over_feature}. In particular, Figure \ref{over_feature}(a), which performs a dimensionality reduction process on the physically observed chemical properties from matminer, which MODNet uses as input features, clearly separates the materials on a planar map with compounds which share either the same working ion or a simlilar crystal system, found in contiguous clusters.

\begin{figure}[h!]
    \centering
    \begin{subfigure}[t]{0.42\textwidth}
        \begin{overpic}[width=\textwidth]{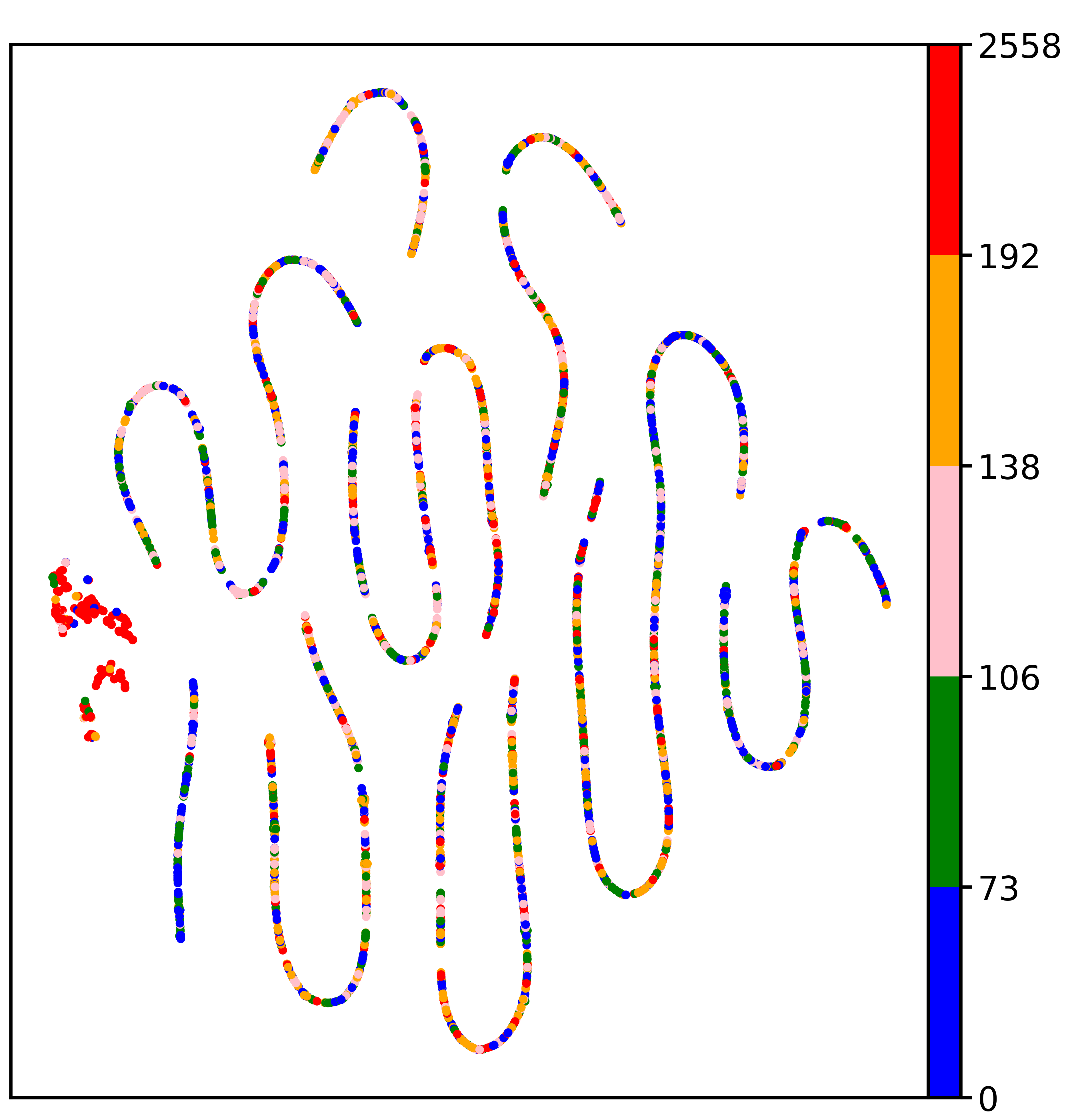}
            \put(0,99){\scriptsize\textbf{(a)}} 
        \end{overpic}
        \label{tsne_modnet_grav}
    \end{subfigure}
    \hspace{-0.3em}
    \begin{subfigure}[t]{0.42\textwidth}
        \begin{overpic}[width=\textwidth]{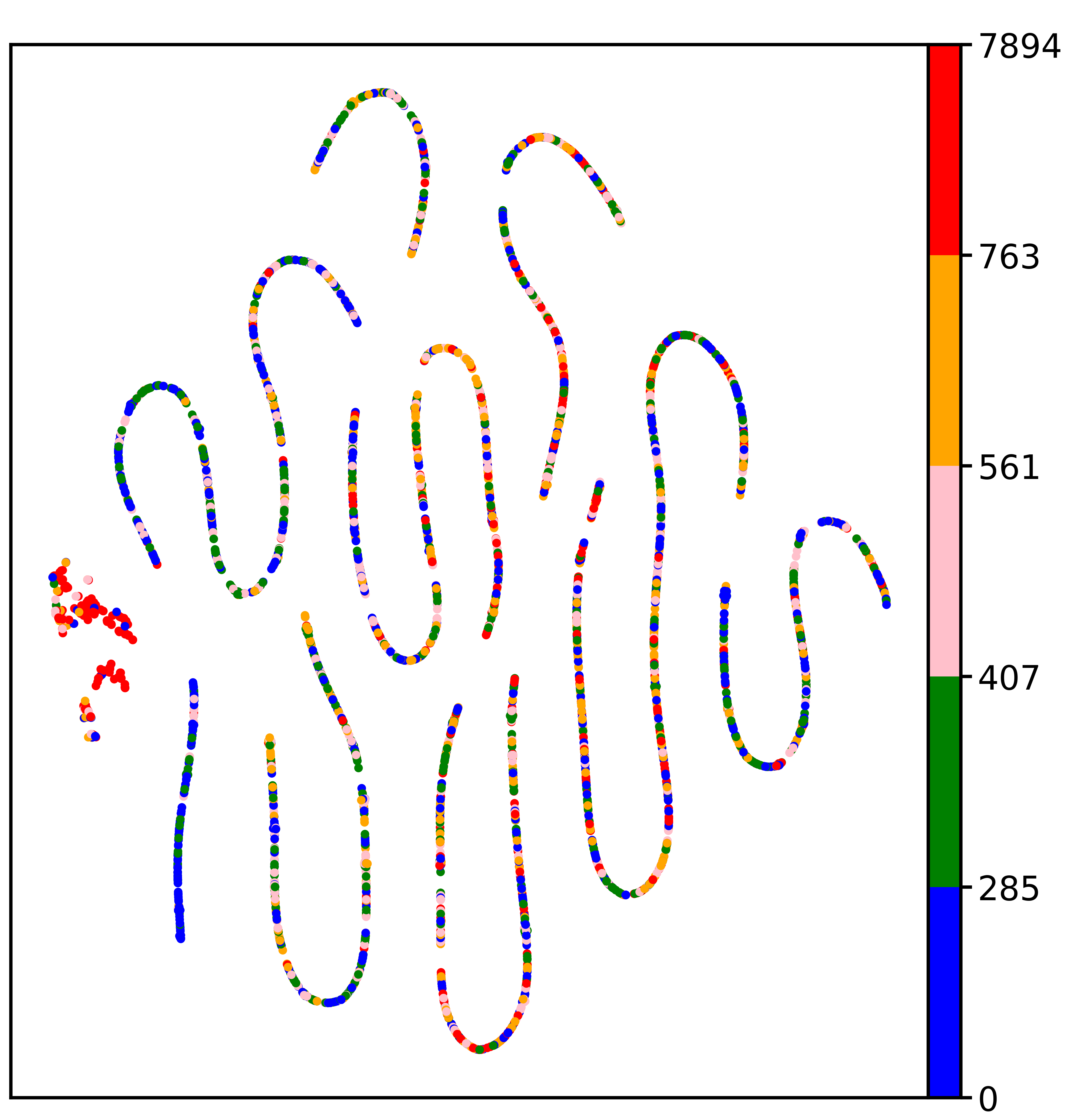}
            \put(0,99){\scriptsize\textbf{(b)}}
        \end{overpic}
        \label{tsne_modnet_vol}
    \end{subfigure}

    \vspace{-0.7em} 

    \begin{subfigure}[t]{0.42\textwidth}
        \begin{overpic}[width=\textwidth]{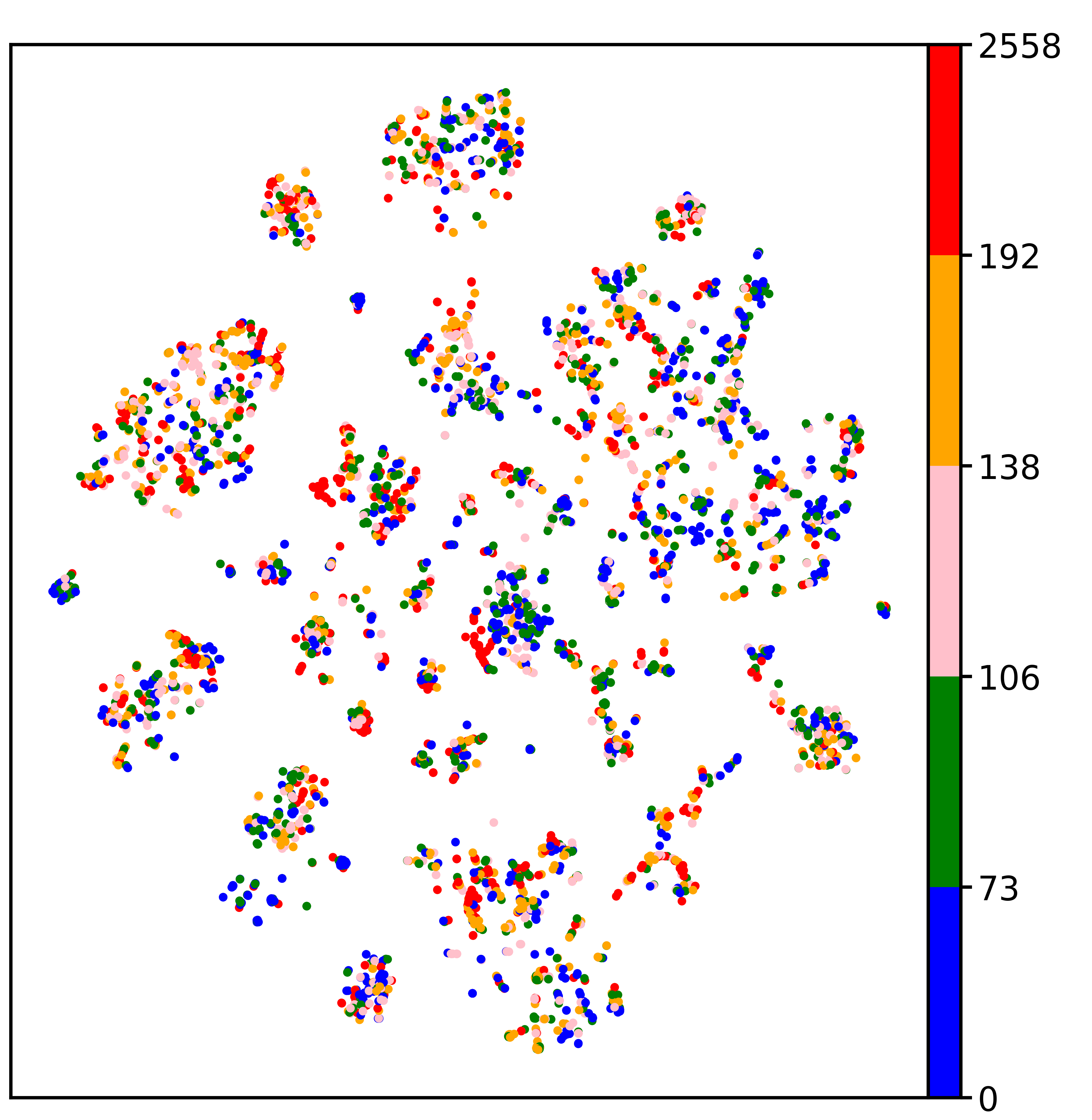}
            \put(0,99){\scriptsize\textbf{(c)}}
        \end{overpic}
        \label{tsne_mat2vec_grav}
    \end{subfigure}
    \hspace{-0.3em}
    \begin{subfigure}[t]{0.42\textwidth}
        \begin{overpic}[width=\textwidth]{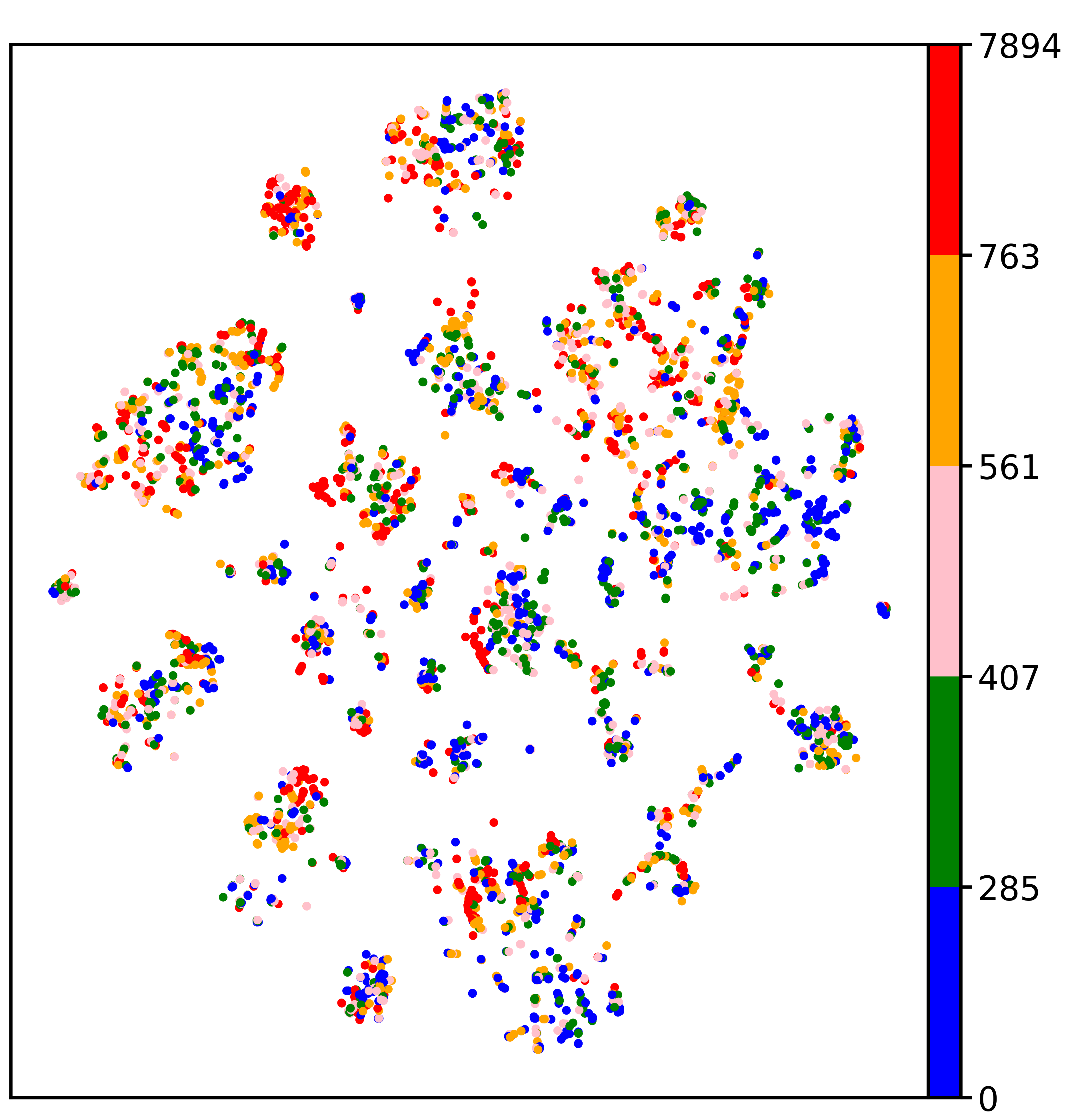}
            \put(0,99){\scriptsize\textbf{(d)}}
        \end{overpic}
        \label{tsne_mat2vec_vol}
    \end{subfigure}

    \caption{2D map of the t-SNE embeddings of the materials using input features from matminer (a, b) and mat2vec (c, d). The points have been colored according to gravimetric (a, c) and volumetric (b, d) capacity quintiles as shown in the provided colorbars.}
    \label{over_capalabel}
\end{figure}

Figure \ref{over_capalabel} shows a 2D-map of the t-SNE embeddings of matminer (MODNet input features) and mat2vec (CrabNet input features) feature vectors. The points have been colored according to their gravimetric and volumetric capacity by equal-frequency binning the target label into five intervals (quintiles). When matminer features are used (panels (a) and (b)), the high gravimetric capacity and high volumetric capacity materials are concentrated in the left-most cluster, distinct from other materials. This highlights the expected relationship between the derived numerical features, a material's chemical composition, and its potential as a cathode material. Subsequent analyses will be carried out using the t-SNE embeddings based on matminer features.

\begin{figure}[h!]
  \centering
  \includegraphics[width=0.7\textwidth]{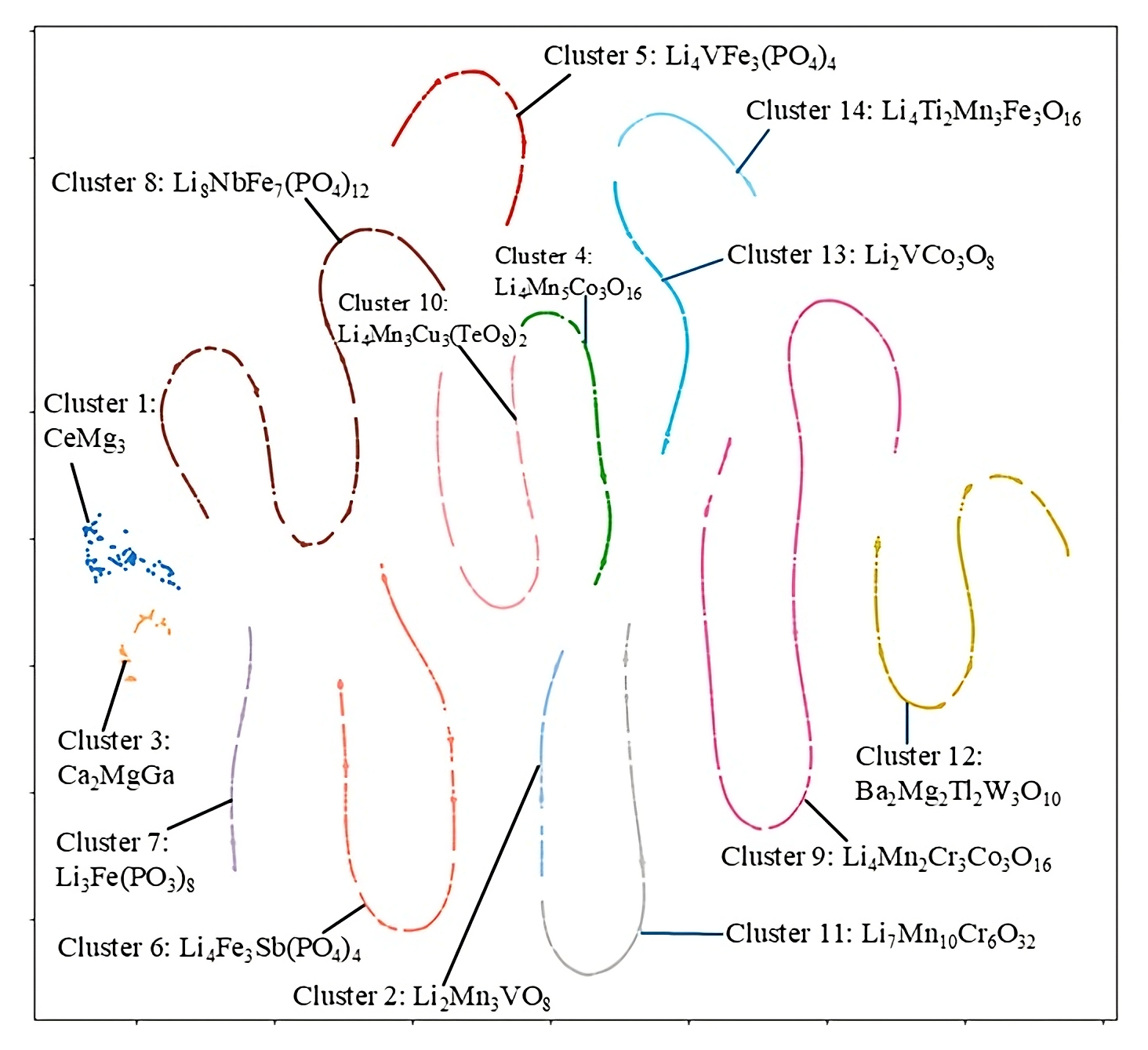}

\caption{2D map of t-SNE embeddings of the materials using input features from MODNet. The points have been colored based on DBSCAN clustering. A total of 14 clusters are identified. The representative material from each cluster, as selected by ElMD mean representative, is indicated together with the cluster number.}
\label{tsne_dbscan}
\end{figure}

To identify groups of chemically similar materials, cluster analysis is performed, based on the embedded points in Figure \ref{over_feature}(a). Cluster labels are automatically assigned to each datapoint using the Density-Based Spatial Clustering of Applications with Noise (DBSCAN) algorithm\cite{khan2014dbscan}, a technique that groups points that are closely packed in the feature space while effectively identifying and excluding outliers from the cluster. By using matminer features, which is to say chemically derived physical properties, we would hope that the embeddings generated with these features would correctly partition the space into chemically distinct regions, which we aim to capture using automated clustering techniques. As illustrated in Figure \ref{tsne_dbscan}, this approach successfully identifies distinct clusters, revealing regions of high local density that correspond to materials with similar chemical and electrochemical characteristics. 

\begin{figure}[h!]
  \centering
  \includegraphics[width=0.99\textwidth]{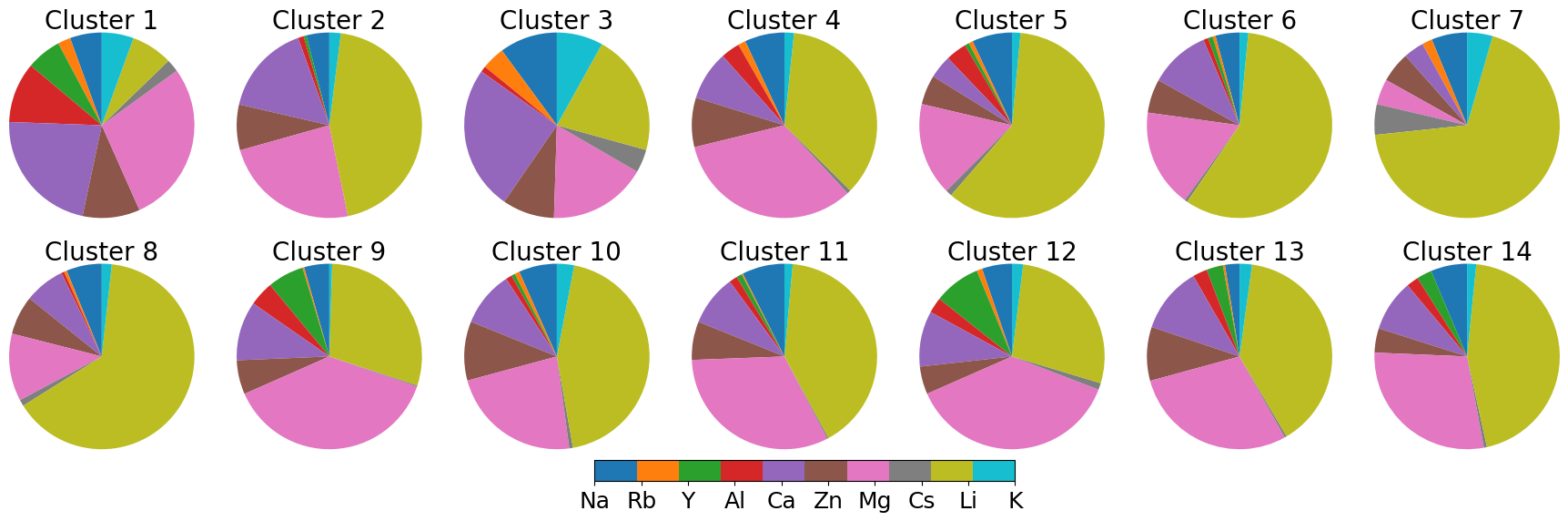}

\caption{Working ion distributions for each of the 14 DBSCAN clusters. }
\label{ion_allclu}
\end{figure}

To summarize the compositional characteristics of the clusters, a barycentric representative material is selected for each of them using the Element Movers Distance (ElMD) ~\cite{hargreaves2020earth}, a metric designed to quantify chemical similarity between materials. ElMD employs the Earth Mover's, or Wasserstein, distance\cite{vallender1974calculation}, an optimal transport-based measure, to compute the minimal ``cost'' required to transform one elemental composition into another. By identifying the material within each cluster that exhibits the lowest average ElMD to all other members of the same cluster, a chemically representative prototype is determined for each label. This approach enables a concise and meaningful interpretation of the typical composition underlying each cluster, allowing for rapid assessment of each clusters overall chemical nature. Whilst it may not precisely uncover the most widely used cathodes as the representative material, it does allow the expert to make a quick assessment that draws on their past experience. For example, lithium iron phosphate, $LiFePO_4$ (LFP), is a widely used cathode material, found in cluster 6 in Figure \ref{tsne_dbscan}. Although this is not a perfect match with the cluster 6 representative material, $Li_4Fe_3Sb(PO_4)_4$, it is evident that this is a doped variant of LFP. The distribution of working ions across all DBSCAN-identified clusters is shown in Figure~\ref{ion_allclu},  revealing how different charge carriers are spatially organized within the embedded feature space. 

To assess the level of distortion introduced by the embedding process, we can compare how well the global geometric structure compares between the high- and low-dimensional points. The global distortion is quantified by comparing average inter-point (Euclidean) distances in the original high-dimensional feature space with those in the embedded low-dimensional representation. These distance matrices are averaged across each DBSCAN cluster for brevity, and reported in Figure \ref{dist_cluster} of the Appendix, which provides a quantitative measure of the structural fidelity retained inside each cluster, and between clusters, during the embedding process. 

\begin{figure}[h!]
  \centering
  \includegraphics[width=1\textwidth]{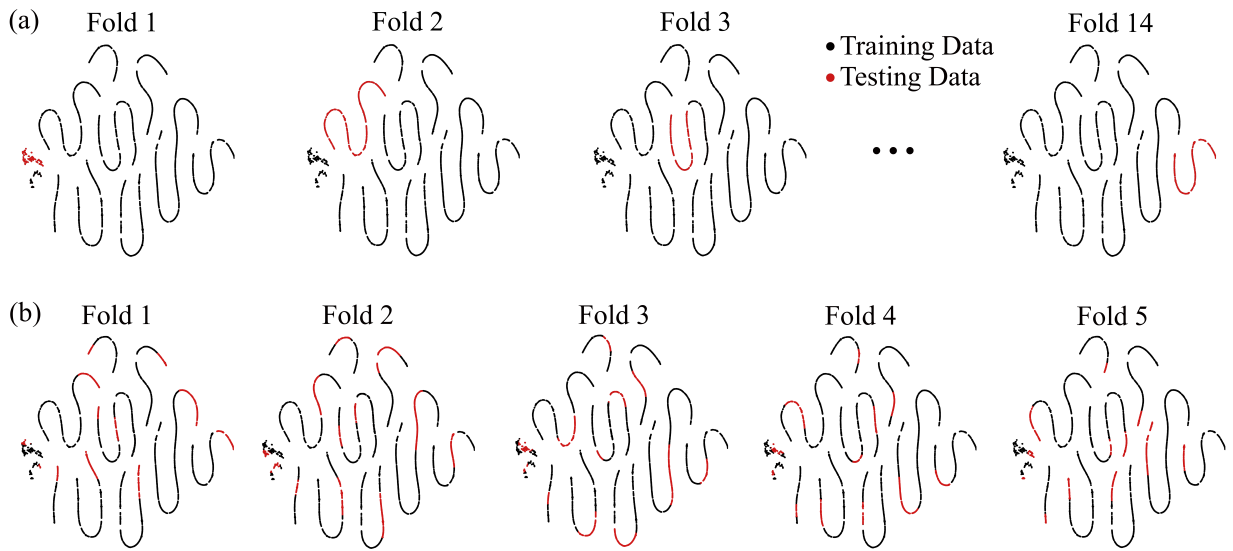}

\caption{(a) Leave One Cluster Out (LOCO) cross-validation, where each of the DBSCAN-assigned clusters is used as a successively held out testing set to assess the model's ability at making out-of-distribution predictions. (b) Stratified \textit{k}-fold cross-validation, where testing sets evenly sample each of the available DBSCAN-assigned clusters for each fold. For visual clarity, the cluster labels in (b) are taken as spatially subsequent subsets from each cluster, rather than being selected randomly. }
\label{tsne_dbscan_labels}
\end{figure}

Cross-validation (CV) is a statistical technique used to quantify a model's ability to generalize to new unseen samples that fall outside its training distribution\cite{allen1974relationship}. As we are always limited by the amount of new data, we must re-use and hold-out our existing data to test how the model might perform on new samples. In \textit{k}-fold cross-validation \cite{stone1978cross}, the dataset is randomly partitioned into \textit{k} mutually exclusive subsets, or folds. During each iteration, one fold is designated as the test set while the remaining \textit{k}-1 folds are used for training, resulting in \textit{k} distinct models being sequentially trained and evaluated. The process is repeated \textit{k} times, with each fold serving exactly once as the test set, ensuring that every data point is used for both training and validation. The final performance metric is computed as the average of the \textit{k} individual scores, providing a robust and statistically sound estimate of model generalization\cite{refaeilzadeh2009cross}. Here, the identified clusters from DBSCAN can be used to implement two other forms of cross-validation to evaluate model performance:
(a) Leave One Cluster Out (LOCO) cross-validation, where the 14 clusters are considered as the folds, with each cluster serving as the test set once, and the remaining clusters being used to train the model.
(b) Stratified 5-fold cross-validation, where each fold is constructed to include one-fifth of a randomly selected sample of the data from every cluster, ensuring that a balanced representation is given across folds, as demonstrated in Figure~\ref{tsne_dbscan_labels}.

The impact of dataset size on model performance is investigated through a bootstrap analysis. Specifically, subsets containing 20\%, 50\%, and 80\% of the original dataset are constructed using uniform sampling. For each subset, unstratified 5-fold cross-validation is employed to determine the average scaled errors across the reported test sets. This analysis serves as a quantitative basis for informing future dataset expansion strategies.

To evaluate and compare the predictive accuracy of the ML models, the scaled mean absolute error (SMAE) \cite{dunn10benchmarking} is utilized, defined as:

$\mathrm{SMAE = \frac{MAE}{MAD} =  \frac{\sum_{i}^{N} |{y_i - {y_i}^{pred}}|}{\sum_{i}^{N} |{y_i - \overline{y}}|}}$

\noindent where MAE represents the mean absolute error, MAD is the mean absolute deviation, $y_i$ is the true label, $y_i^{\text{pred}}$ is the predicted label, and $\overline{y}$ is the average value of the true labels. The SMAE normalizes the MAE by the inherent variability of the target property, thereby enabling a statistically fair comparison across different electrochemical properties with varying scales. 

As a control, a mean predictor (null) model was implemented, which assigns the arithmetic mean of the training-set labels to all samples in the test set. The control baseline provides a theoretical maximum expected error, representing a scenario in which the model fails to capture any underlying chemical relationships and instead only learns the average of the training data target label. This serves as a numeric proxy for the real-world random guessing scenario. In this study, all evaluated ML models significantly outperform this baseline, giving us confidence in their use as ranking models for high-throughput screening.

\section*{Results \& Discussion}

The SMAE results are summarized in Table \ref{tab:results}. Of the three evaluated models, from these metrics, CrabNet consistently delivers better predictive performance on these target labels. 

In order to compare our results to previous works, we also adopt the methodology of Ref. \cite{zhang2024interpretable}, whereby outliers falling outside a $2\sigma$-range of the target values are removed. The testing errors for this filtered 2$\sigma$-datasets are reported in Table \ref{tab:results} (brackets). When predicting gravimetric capacity, CrabNet, operating solely on composition-based features, outperforms Extreme Tree Regression (ETR) (MAE = $24.482$) and achieves accuracy comparable to the Light Gradient Boosting Machine (LGBM) and Deep Neural Network (DNN) models reported in a previous work \cite{zhang2024interpretable}, despite those models requiring structural descriptors. This performance again demonstrates the utility of composition-based approaches, which circumvent the need for computationally expensive structural data, as an effective first-stage screening tool for high-throughput materials discovery.

\begin{table}[h!]
\centering
\caption{MAE and SMAE on the target properties for the different models (MODNet, Crabnet, and RF@Magpie) from 5-fold cross-validation, for the complete dataset and the filtered $2\sigma$-datasets in the brackets. The best result for each column is highlighted in bold. The control model assigns the arithmetic mean of the training-set labels to all samples.}
\renewcommand{\arraystretch}{1.5}
\setlength{\tabcolsep}{4pt}

\newcolumntype{C}[1]{>{\centering\arraybackslash}p{#1}}
\resizebox{\textwidth}{!}{
\begin{tabular}{C{0.142\textwidth}
C{0.094\textwidth}
C{0.094\textwidth}
C{0.094\textwidth}
|
C{0.094\textwidth}
C{0.094\textwidth}
C{0.094\textwidth}
|
C{0.094\textwidth}
C{0.094\textwidth}
C{0.094\textwidth}}
\hline\hline
\multirow{2}{*}{Model} 
 & \multicolumn{3}{c}{Gravimetric capacity (mAh/g)} 
 & \multicolumn{3}{c}{Volumetric capacity (mAh/cm$^3$)} 
 & \multicolumn{3}{c}{Average voltage (V)} \\
\rule{0pt}{2.6ex}
 & MAE & SMAE & $R^2$
 & MAE & SMAE & $R^2$
 & MAE & SMAE & $R^2$ \\
\hline
MODNet      & 26.834 (21.085) & 0.308 (0.242) & 0.841 (0.726) & 106.252 (86.173) & 0.333 (0.270) & 0.810 (0.739) & 1.129 (0.634) & 0.489 (0.277) & 0.051 (0.699) \\
CrabNet     & \textbf{24.730 (18.126)} & \textbf{0.284 (0.208)} & \textbf{0.843 (0.724)} & \textbf{94.312 (77.805)} & \textbf{0.295 (0.244)} & \textbf{0.830 (0.722)} & \textbf{1.087 (0.653)} & \textbf{0.474 (0.285)} & \textbf{0.090 (0.660)} \\
RF@Magpie   & 49.180 (35.166) & 0.565 (0.404) & 0.643 (0.533) & 173.328 (137.967) & 0.543 (0.432) & 0.646 (0.540) & 1.588 (0.925) & 0.693 (0.404) & 0.084 (0.562)\\
Control     & 87.095 (27.739) & 1.000  (0.663) & 0.000 (0.000) & 319.238 (232.880) & 1.000  (0.729) & 0.000 (0.000) & 2.292 (1.858) & 1.000  (0.811) & 0.000 (0.000) \\
\hline\hline
\end{tabular}
}
\label{tab:results}
\end{table}

One observation that can be seen across all models is the drop in predictive accuracy between the gravimetric capacity and the average voltage. While the capacity, a stoichiometric property fundamentally determined by the number of redox-active sites and the molar mass, is effectively captured by compositional features, predicting the average voltage is more challenging to resolve. From a thermodynamic perspective, the average voltage is governed by the difference in chemical potential of the working ion between the host electrode and the anode. This property is sensitive to the local coordination environment and the inductive effects of the anion framework. For instance, the voltage of a redox couple (e.g., $Fe^{2+}/Fe^{3+}$) will change when transitioning from an oxide to a polyanionic framework due to changes in the ionicity of the metal-ligand bond, a structural nuance that purely composition-only models can only partially approximate without explicit site-symmetry or bond-length data. Nevertheless, despite these inherent physical limitations, we believe our composition-based framework remains highly valuable for the initial stages of the materials discovery pipeline. By bypassing the need for computationally expensive and often unavailable structural characterization, these models enable high-throughput screening of a broader expanse of candidate materials space. Thus, rapid screening of vast numbers of compositions may be carried forward to whittle the list down from billions of potential materials to a more manageable subset, which can then be refined using high-fidelity structural simulations or experimental synthesis. In this way, our approach serves as an essential, computationally efficient filter that can significantly accelerate the early-stage identification of high-performance electrode materials.

\begin{figure}[!h]
    \centering
    \makebox[\textwidth][c]{\scriptsize\textbf{Gravimetric capacity}\par}%
    \vspace{0.7em}
    \begin{subfigure}{0.4\textwidth}
        \begin{overpic}[width=\textwidth,trim=0 4 0 0,clip]{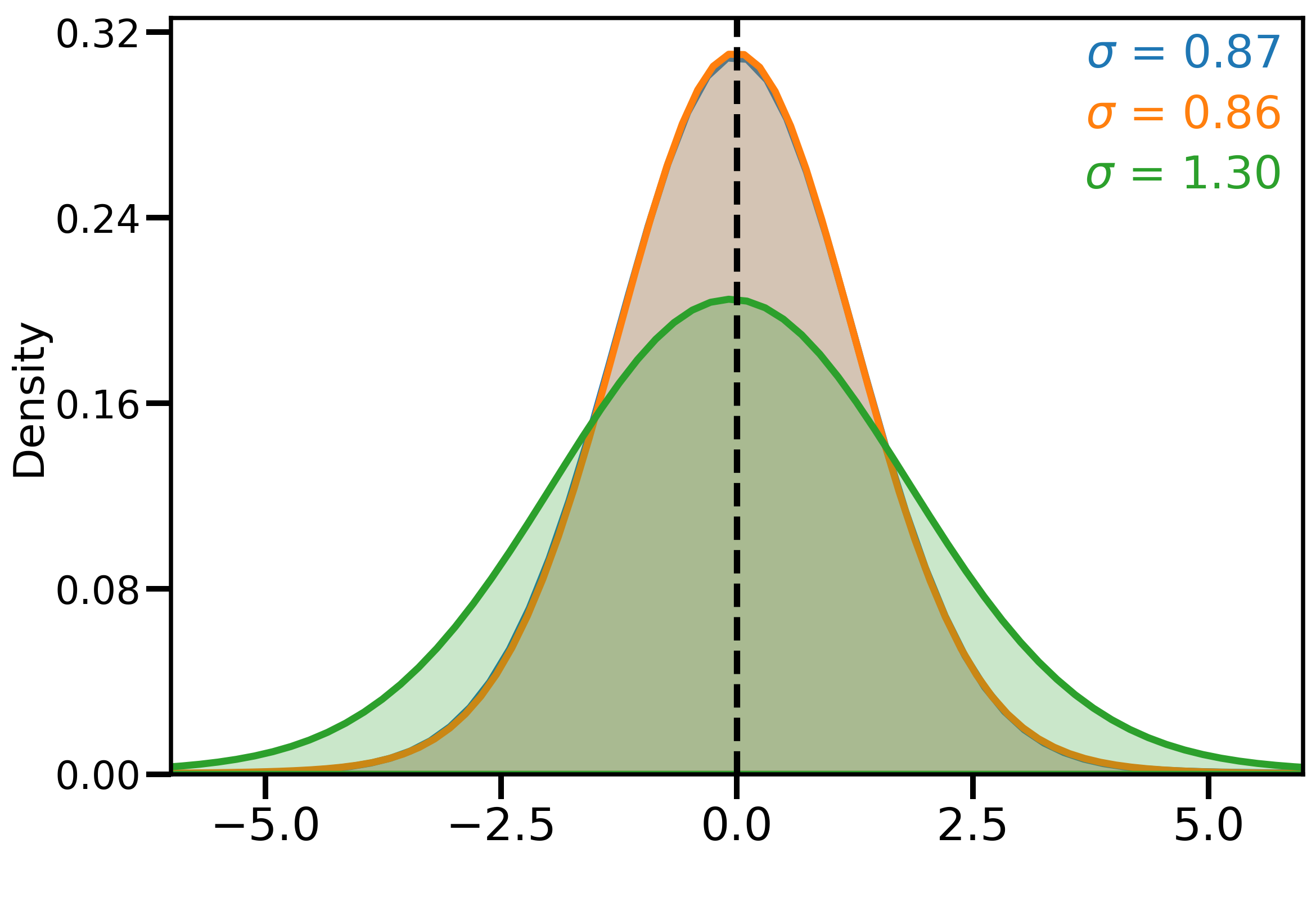}
            \put(2.5,67){\scriptsize\textbf{(a)}}
        \end{overpic}
    \end{subfigure}
    \hspace{-0.1em}
    \begin{subfigure}{0.4\textwidth}
        \begin{overpic}[width=\textwidth,trim=0 4 0 0,clip]{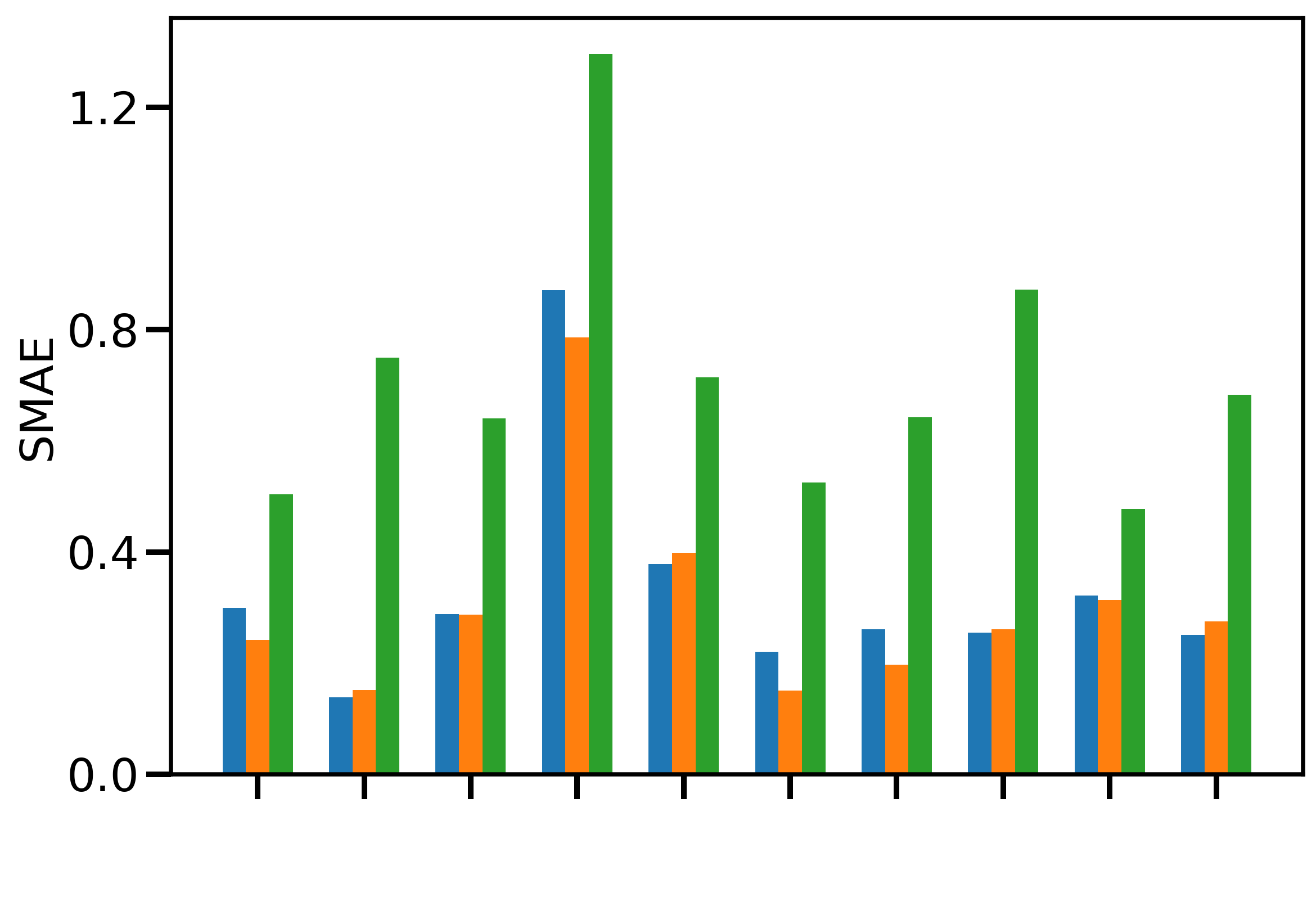}
            \put(2.5,67){\scriptsize\textbf{(b)}}
        \end{overpic}
    \end{subfigure}

    \vspace{-1.3em}
     
    \makebox[\textwidth][c]{\scriptsize\textbf{Volumetric capacity}\par}%
    \vspace{0.7em}

    \begin{subfigure}{0.4\textwidth}
        \begin{overpic}[width=\textwidth,trim=0 4 0 0,clip]{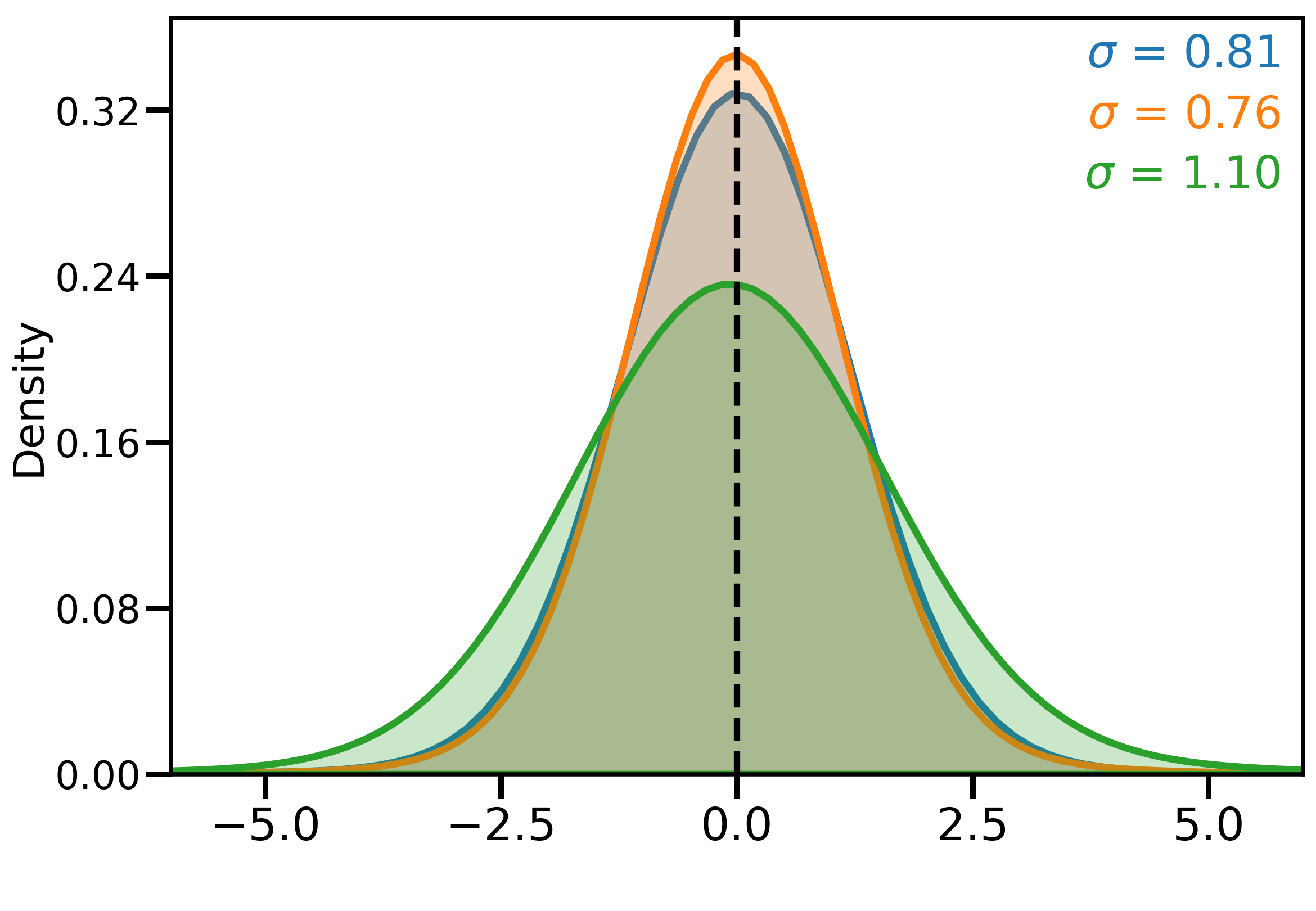}
            \put(2.5,67){\scriptsize\textbf{(c)}}
        \end{overpic}
    \end{subfigure}
    \hspace{-0.1em}
    \begin{subfigure}{0.4\textwidth}
        \begin{overpic}[width=\textwidth,trim=0 4 0 0,clip]{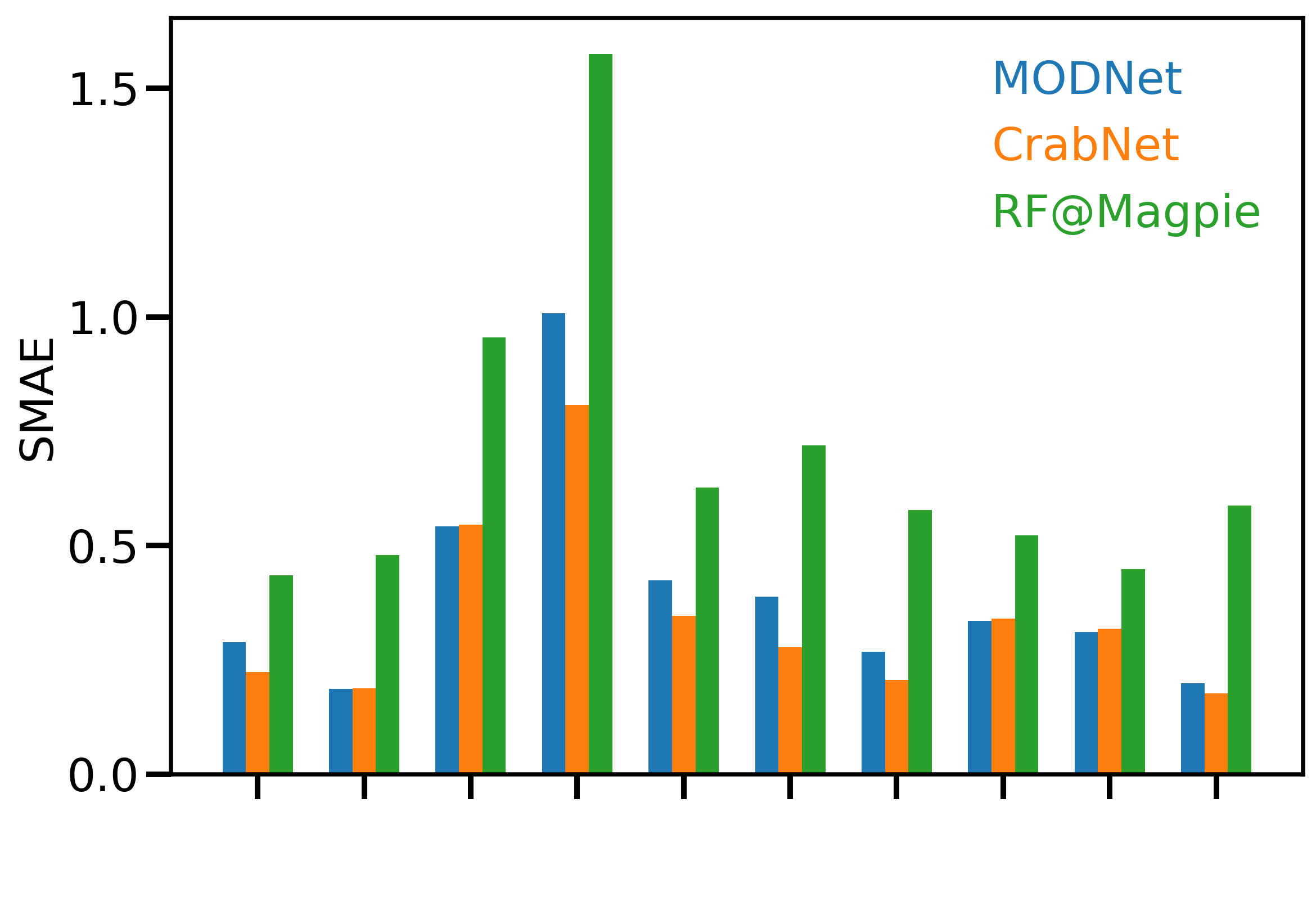}
            \put(2.5,67){\scriptsize\textbf{(d)}}
        \end{overpic}
    \end{subfigure}

    \vspace{-1.3em}
    
    \makebox[\textwidth][c]{\scriptsize\textbf{Average voltage}\par}%
    \vspace{0.7em}

    \begin{subfigure}{0.4\textwidth}
        \begin{overpic}[width=\textwidth]{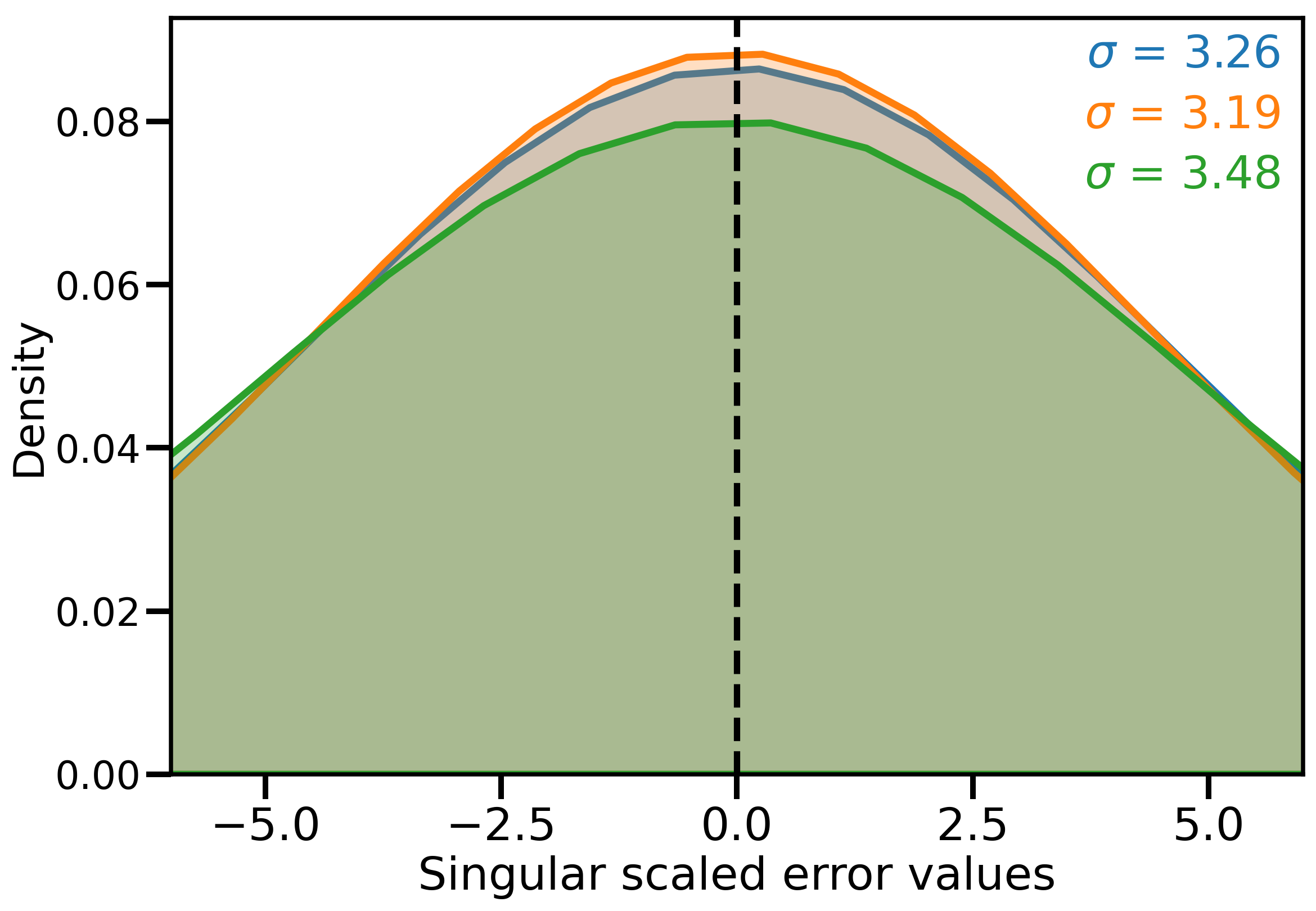}
            \put(2.5,70){\scriptsize\textbf{(e)}}
        \end{overpic}
    \end{subfigure}
    \hspace{-0.1em}
    \begin{subfigure}{0.4\textwidth}
        \begin{overpic}[width=\textwidth]{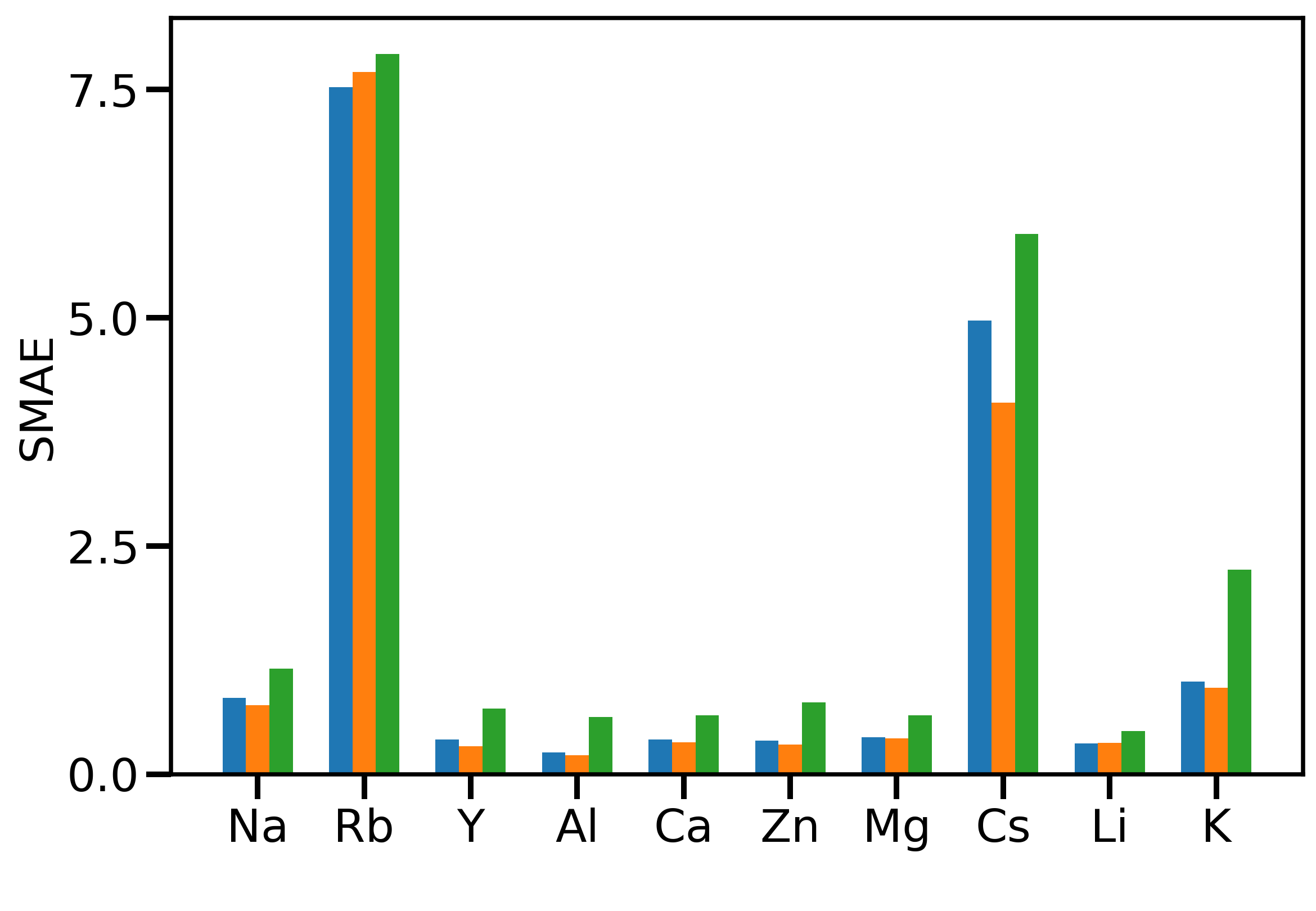}
            \put(2.5,70){\scriptsize\textbf{(f)}}
        \end{overpic}
    \end{subfigure}

    \caption{Singular scaled error values (singular testing error values scaled by the same MAD as the calculation in SMAE) distributions on the target properties for the different models (MODNet in blue, Crabnet in orange, and RF@Magpie in green). For panels (a), (c), and (e), the kernel density estimation (KDE) plot provides an estimation of the testing error values distribution, using a Gaussian kernel. Panels (b), (d), and (f) represent the view of the SMAE values (y axis) for each of the electrode materials by working ion. Panels (a) and (b) show the gravimetric capacity, (c) and (d) the volumetric capacity, (e) and (f) the average voltage.}
    \label{kde_grav_vol_ion}
\end{figure}

Figure \ref{kde_grav_vol_ion} illustrates the distribution of testing errors for the three models under unstratified 5-fold cross-validation, estimated via a Gaussian kernel density function. The CrabNet error distributions are more densely concentrated near zero, with superior predictive accuracy. By contrast, RF@Magpie consistently yields the highest prediction errors across each of the investigated properties.

The average SMAE, partitioned by the working ions of the electrode materials, is presented in Figures \ref{kde_grav_vol_ion}(b), (d), and (f). Notably, the largest sets with more samples, such as Li (43.6\% MP battery dataset) and Mg (25.6\% MP battery dataset), have a lower SMAE. In general, the SMAE of the model correlates with the number of training samples. Despite this trend, the lowest predictive accuracy for gravimetric and volumetric capacities is for the aluminum-based materials, and is significantly lower than the other working ions. Similarly, materials utilizing rubidium and cesium as working ions display a significant performance degradation in predicting average voltage. In addition to the reduced number of samples, this discrepancy may be attributable to the featurization stage; because these elements appear infrequently in upstream training data, the statistical alignment between provided numeric representations and their specific chemical properties may be underdeveloped.

\begin{figure}[h!]
\centering
  \includegraphics[width=0.8\textwidth]{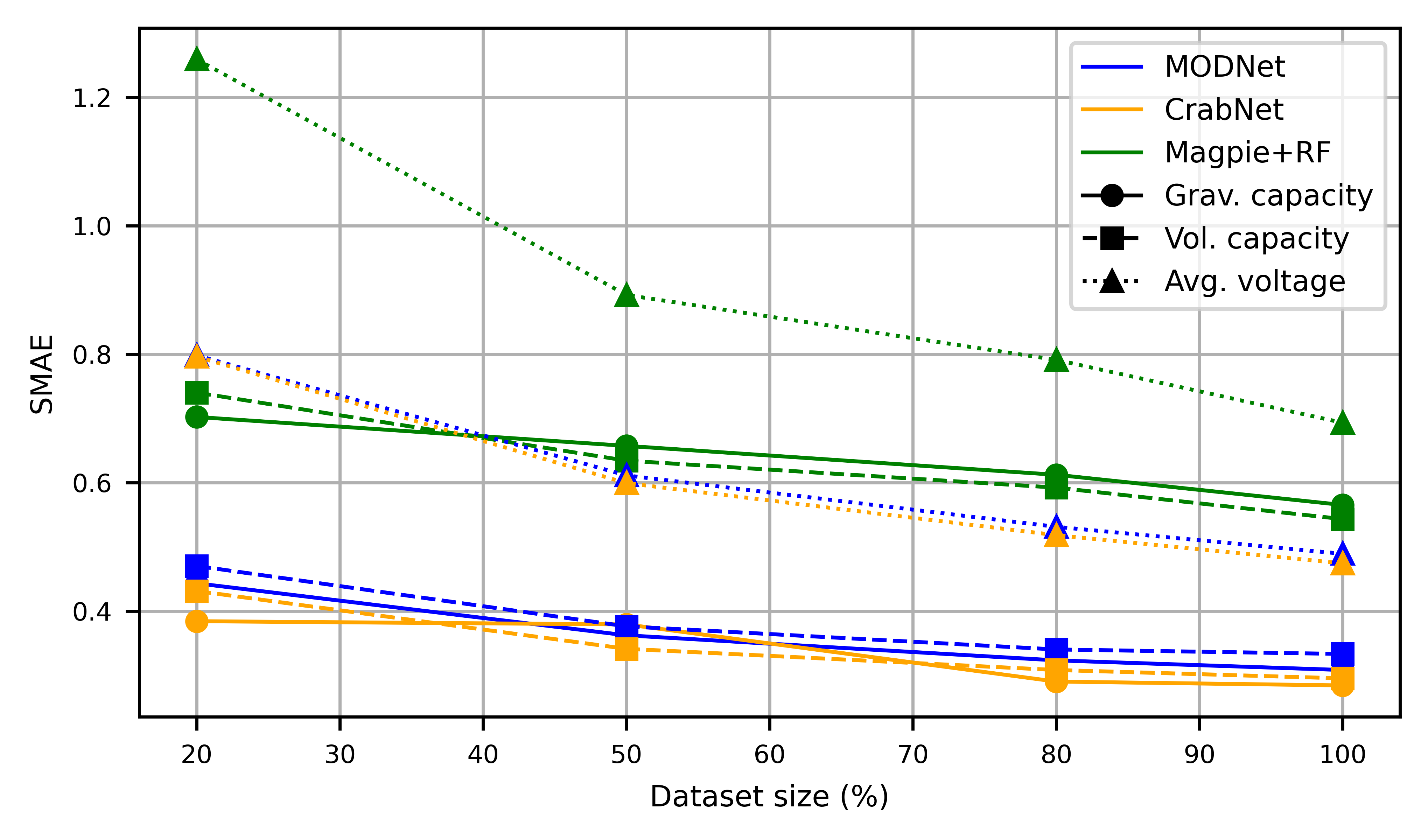}\hfill
  
\caption{SMAE on the target properties for the different models (MODNet, Crabnet, and RF@Magpie) as a function of the dataset size (expressed in \% of the total size) for each of the three for each properties,}
\label{err_percen}
\end{figure}

The results of the bootstrap analysis are shown in Figure \ref{err_percen}. As expected, predictive errors decrease monotonically as dataset size increases. This trend highlights the critical importance of expanding large-scale materials datasets to enhance the performance and reliability of machine learning models in materials discovery.

Table \ref{tab:overall_spcv} reports the SMAE scores for Leave-One-Cluster-Out (LOCO) and Stratified $k$-fold (SkF) cross-validation, normalized using the same MAD as in Table \ref{tab:results}. These schemes assess model robustness under distinct partitioning strategies: LOCO simulates out-of-distribution scenarios, while SkF ensures balanced sampling across all identified clusters. Errors from LOCO CV are consistently higher than those from stratified 5-fold CV. This trend is expected, as LOCO forces the test set to be chemically distinct from the training distribution, whereas SkF includes representative samples from every cluster in both the training and testing sets. In addition, SkF errors are obviously higher than those in Table \ref{tab:results}, which means that the uniformly distributed train-test folds of all groups failed to achieve a better representative compared to the unstratified case, suggesting the limitation of attaching the clustering results with chemical features.

\begin{table}[h!]
\centering
\caption{SMAE on the target properties for the different models (MODNet, CrabNet, and RF@Magpie) obtained through Leave One Cluster Out (LOCO) and Stratified \textit{k}-fold (SkF) cross-validation. The best result for each column is highlighted in bold. The control model assigns the arithmetic mean of the training-set labels to all samples.}
\setlength{\tabcolsep}{4pt}

\newcolumntype{C}[1]{>{\centering\arraybackslash}p{#1}}
\begin{tabular}{
C{0.142\textwidth}
C{0.094\textwidth}
C{0.094\textwidth}
|
C{0.094\textwidth}
C{0.094\textwidth}
|
C{0.094\textwidth}
C{0.094\textwidth}
}
\hline\hline
\multirow{2}{*}{Model} 
 & \multicolumn{2}{c}{Gravimetric capacity (mAh/g)} 
 & \multicolumn{2}{c}{Volumetric capacity (mAh/cm$^3$)} 
 & \multicolumn{2}{c}{Average voltage (V)} \\
\rule{0pt}{2.6ex}
 & LOCO & SkF 
 & LOCO & SkF 
 & LOCO & SkF \\
\hline
MODNet & 0.652 & 0.448 & 0.586 & 0.466 & 0.663 & 0.574 \\
CrabNet & \textbf{0.551} & \textbf{0.388} & \textbf{0.530} & \textbf{0.436} & \textbf{0.603} & \textbf{0.562} \\
RF@Magpie & 1.028 & 0.717 & 0.933 & 0.723 & 1.113 & 0.804 \\
Control & 1.154 & 0.957 & 1.145 & 0.930 & 0.947 & 0.910 \\
\hline\hline
\end{tabular}%

\label{tab:overall_spcv}
\end{table}

Comparing the LOCO and SkF results in Table \ref{tab:overall_spcv}, \ref{tab:spcv1re}, and \ref{tab:spcv2re} with the unstratified 5-fold CV scores in Table \ref{tab:results} displays the same trend as before, with CrabNet achieving superior predictive performance across all schemes. However, testing errors vary substantially across individual clusters. For instance, clusters 1 and 3 exhibit relatively high errors, while clusters 4, 7, and 9 yield significantly lower errors.

Referring to Figure \ref{tsne_dbscan} and Figure \ref{dist_cluster}, clusters 1 and 3 are positioned near the periphery of the 2D t-SNE projection in isolated regions, which suggests that their MODNet-derived features may differ substantially from those of other materials. This separation may reflect distinct chemical compositions that challenge model generalization and contribute to the elevated testing errors observed for these specific clusters. In contrast, the remaining clusters are more centrally located, exhibit lower error values, and display clearer partitioning relative to clusters 1 and 3. Referring back to Figure \ref{ion_allclu}, these two clusters have a lower distribution of Li than clusters 2 and 4-14, and as these clusters benefit from larger sample sizes, this could explain their lower testing errors.

The embedding process is an approximate representation of the space, which must distort the high dimensional structure. To numerically analyze the resultant cluster structure, we compute a full pairwise Euclidean distance matrix for the embedded points, which is averaged across each cluster (Figure \ref{dist_cluster}, where the average inter-point distance of each cluster is provided along the anti-diagonal). This gives a quantitative measure of the intra- and inter-cluster similarity in 2D t-SNE space. For comparative analysis, we also compute a corresponding distance matrix in the original high-dimensional feature space, which allows for an assessment of the level of distortion between the original and embedded representations. Figure \ref{dist_cluster} shows that the intra-cluster anti-diagonal entries, are generally smaller than distances between different clusters, as materials belonging to the same cluster exhibit greater similarity in both the original feature space and the 2D t-SNE embedding. At the same time, the spread of these intra-cluster distances demonstrates that the embedded points express a level of heterogeneity within each cluster, with some degree of the original structural variation preserved in the embedding. 

The relationship between these two matrices can be demonstrated as a scatter plot, by plotting each of the average distance across each cluster in high-dimensional space and t-SNE embedded space as the $x$ and $y$ intercepts for each point, Figure \ref{corr_dist}. 
Figure \ref{corr_dist}(a) again reveals two distinct regions (colored by blue and red), separated at a normalized original-space distance of $x\approx0.8$. When examined alongside Figure \ref{dist_cluster}, distances exceeding this threshold are primarily associated with pairwise relationships involving clusters 1 and 3 (red), whereas smaller distances correspond to relationships among the remaining clusters (blue). 

For all distance pairs between different clusters with $x<0.8$, we compute both the distance correlation \cite{szekely2007measuring}, which captures nonlinear dependence, and the normalized mutual information (NMI) between the original-space and embedded-space distances. The resulting distance correlation of 0.45 indicates a moderate nonlinear association between the two representations, while the high NMI value of 0.81 suggests high similarity in the original feature space. Together, these metrics suggest that the global separation patterns and local distance relationships from the original space are somewhat preserved in the two-dimensional t-SNE embedding. To further examine geometric consistency, a linear regression is performed for the $x<0.8$ inter-cluster subset of points from Figure \ref{corr_dist}(a), shown in Figure \ref{corr_dist}(b), yielding an $R^2$ value of 0.22. This result indicates that the linear relationships associated with the cluster-level geometry is only partially maintained under t-SNE.

The positive slope and statistically meaningful correlations confirm that relative cluster separations are retained, whereas the limited explanatory power of the linear model underscores the intrinsically nonlinear nature of t-SNE and its emphasis on preserving local neighborhood structure rather than global metric fidelity. Overall, t-SNE provides a qualitatively faithful representation of the clustering structure, maintaining major separations and boundaries, as reflected by the strong NMI, while introducing nonlinear distortions in absolute distances, as evidenced by the regression analysis.
 
In addition to the LOCO-validation, further testing errors are obtained via stratified 5-fold cross-validation, which are reported in Table \ref{tab:spcv2re}. The ranked performance of the models remains consistent with Table \ref{tab:results}, again placing CrabNet as the most accurate predictor. However, while the overall ranking remains unchanged, the performance of the RF@Magpie model declines markedly under this specific validation scheme. As it consistently displays better performance than other models, for our final model, we use a CrabNet architecture trained on the complete MP battery dataset.  

Due to the final models intended use as a practical screening tool, we further validate the final model using unseen experimentally reported data. This introduces a new challenge, as there are no large datasets of experimentally validated cathode compositions with their measured properties, and the MP battery dataset already contains many of the reported compounds. In this study, to assess the practical generalizability and limitations of the trained final model, we have collated a new cathode dataset based on an entirely unseen cohort of seven experimentally validated materials; taking or calculating the required properties from the given values and plots in the report. 

The materials belong to several different crystal classes, such as dense, high-capacity layered oxides, such as $\text{Li}_{1.2}\text{Ni}_{0.2}\text{Mn}_{0.6}\text{O}_2$, and $\text{LiNi}_{0.33}\text{Mn}_{0.33}\text{Co}_{0.33}\text{O}_2$, to lower-density open-framework tunnel structures like $\text{Na}_{0.44}\text{MnO}_2$, and the polyanionic fluorophosphate $\text{Na}_3\text{V}_2(\text{PO}_4)_2\text{F}_3$. Measuring the performance of the final model against this data provides us with a quantified measure of reliability for how we could expect the final model to perform when screening new compositions in real-world scenarios. 

\begin{figure}[h!]
    \centering
    \begin{subfigure}[t]{0.33\textwidth}
        \begin{overpic}[width=\textwidth]{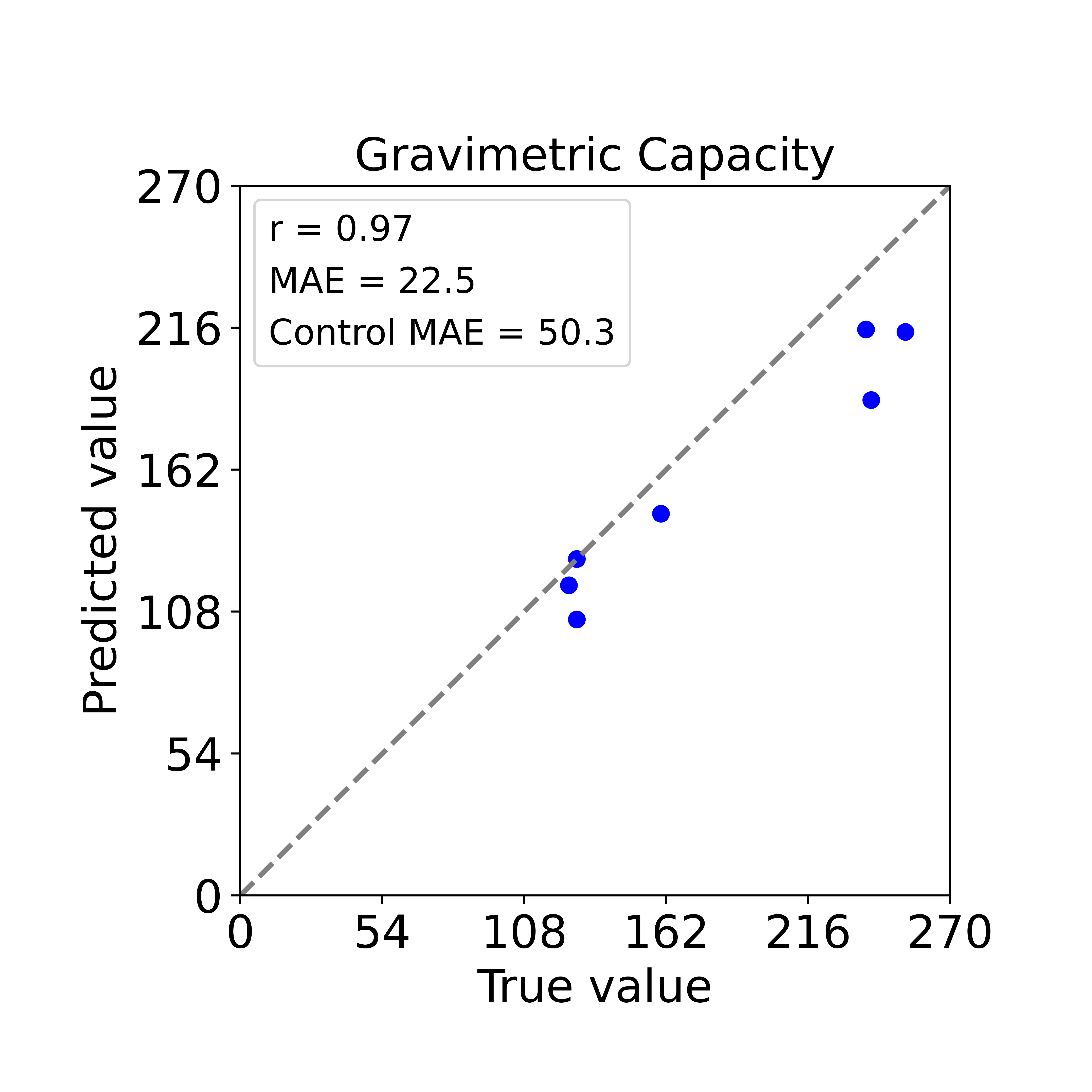}
            \put(5,88){\scriptsize\textbf{(a)}} 
        \end{overpic}
        \label{exp_pred_grav}
    \end{subfigure}
    \hspace{-1.5em}
    \begin{subfigure}[t]{0.33\textwidth}
        \begin{overpic}[width=\textwidth]{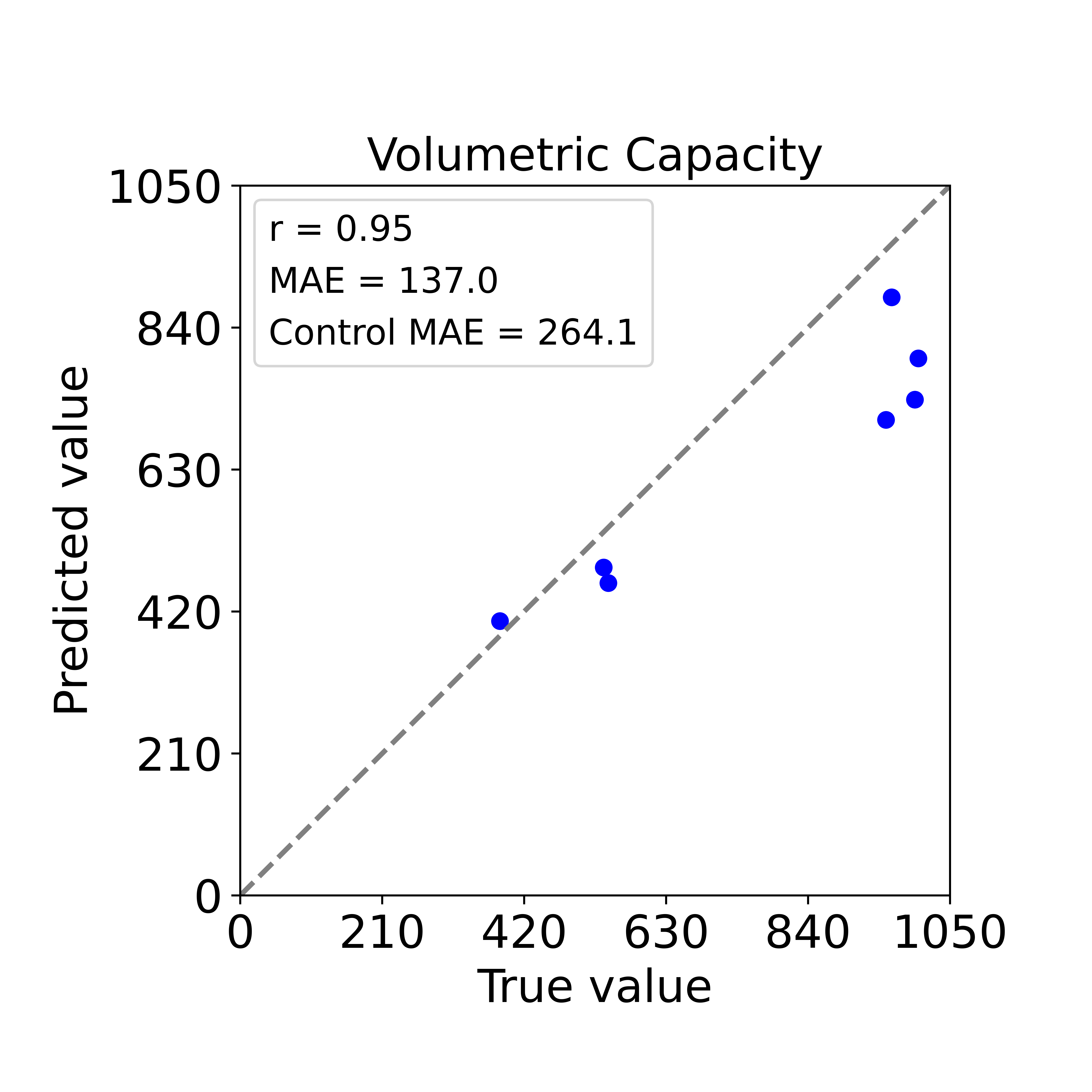}
            \put(5,88){\scriptsize\textbf{(b)}}
        \end{overpic}
        \label{exp_pred_vol}
    \end{subfigure}
    \hspace{-1.5em}
    \begin{subfigure}[t]{0.33\textwidth}
        \begin{overpic}[width=\textwidth]{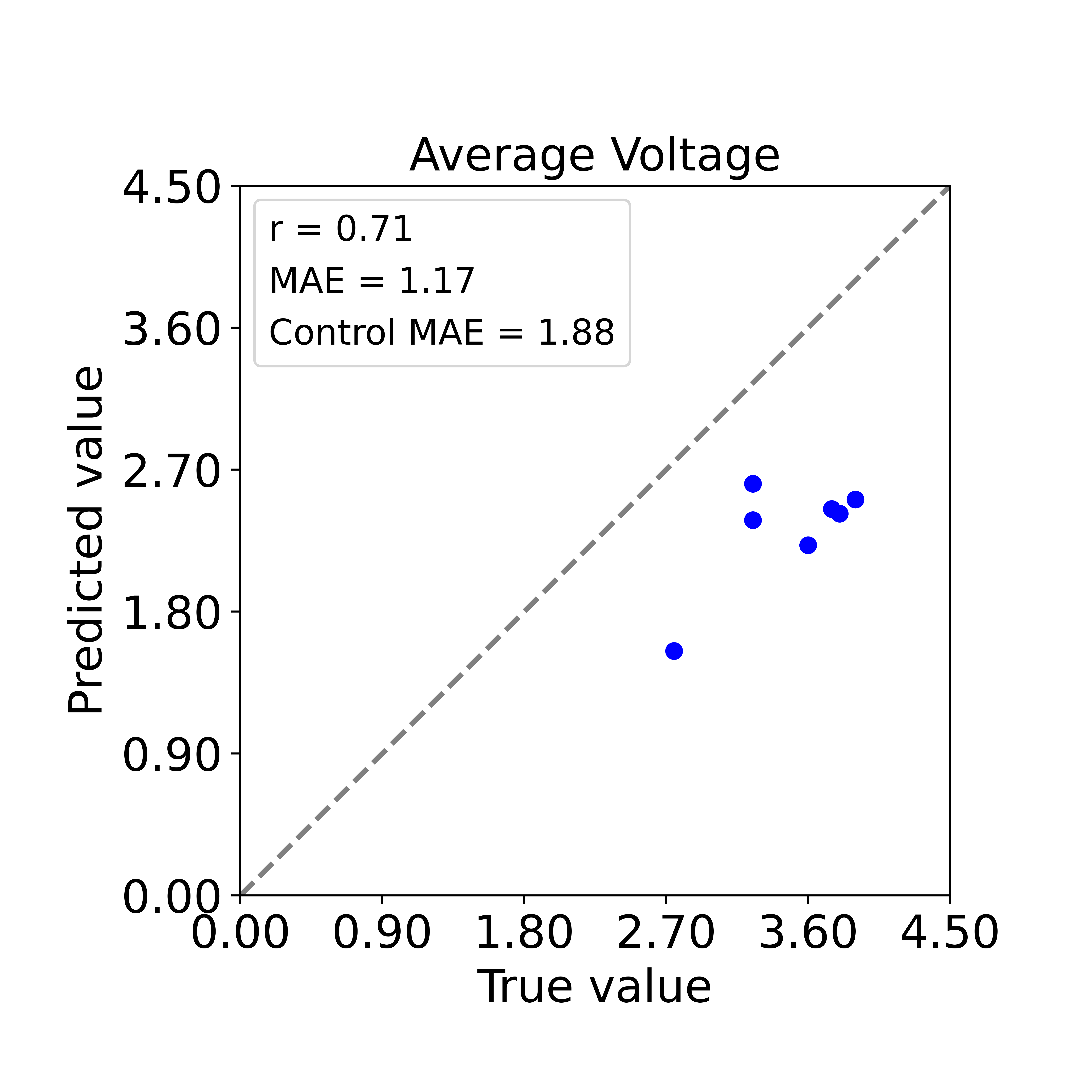}
            \put(5,88){\scriptsize\textbf{(c)}}
        \end{overpic}
        \label{exp_pred_volt}
    \end{subfigure}
    \caption{Parity plots demonstrating the predicted (y-axis) and true (x-axis) values showing how each of the experimentally validated materials in the hold-out set perform (blue) for each of the target properties. The predictions made by the ideal model, where each of the scores would fall on the $x=y$ regression line, is shown in dashed grey. The Pearson's $r$-score is provided in each of the legends, along with the Mean Absolute Error (MAE) for each of the seven points. In addition, the MAE of the three control models is given to directly compare model performance against a real-world baseline. As these controls simply predict the average of the training values, by definition they will have an undefined $r$-score.}
    \label{exp_pred}
\end{figure}

We test the final trained CrabNet model against this experimental dataset of cathode materials \cite{lee2004synthetic, marker2019evolution, hong2010structural, acs2026tailored, cao2011reversible, pamidi2023single, camacho2025lattice}, which is provided in the Appendix, Table \ref{tab:expvali}. The parity plots (true vs. predicted values) are shown in Figure \ref{exp_pred}, where each set of points display a positive trend, which is reflected by their strong to reasonable Pearson's correlation $r$-scores. Further, by re-using the same control models from the previous section, which reflect the ability of a very poor model that has only learnt the mean of labels of the training set, we can again compare how the final CrabNet model improves against the baseline of predictions numerically. For the gravimetric capacity, despite a consistent bias to under-predict the values, these are strongly correlated with the true labels. This is further reflected by the improvement in MAE against the control, from $50.3$ to $22.5 \text{ mAh/g}$.

For the volumetric capacity, the CrabNet model also provides predictions with a strong correlation between the predicted and true values, $r=0.95$. Similar to before, when compared to the control, the MAE improves from $264.1$ to $137.0$. As discussed in previous sections, the calculated average voltage labels which have been provided in the training dataset are suspicious, and we would not expect the model to perform particularly well when provided with only the composition of the material, with no structural information. Nevertheless, the models still attain a positive correlation, although the results clearly possess greater bias and poorer resolution than the other two properties, which is reflected by it's $r$-score of $0.71$ and a weaker improvement against the control MAE from $1.88$ to $1.17$.

The computational dataset assumes idealized 0K structures, in perfect infinite arrangements of atoms, undergoing complete, frictionless multi-electron transitions. However, the performance of electrode materials is altered through the realities introduced by the crystal synthesis process, such as reduced ion-mobility via grain-boundary transitions, inter-facial polarizations, or localized kinetic phase-trapping. 

Despite these challenges, the final trained model preserves the general rankings of each of the materials. In practical terms, this absolute numerical offset does not diminish the model's utility as an accelerated pre-screening tool, where the desired output is a ranking of a list of candidate compounds. The further integration of structural descriptors may be required to overcome the systematic voltage shifts and attain accurate predictions for chemical descriptors, however, from the presented results, we believe the current architecture is effective for high-throughput down-selection at the compositional stage.

\clearpage

\section*{Conclusion}

This comparative study evaluates leading compositional models for predicting electrode properties from chemical formulations, specifically assessing the performance of MODNet, CrabNet, and RF@Magpie on the Materials Project Battery dataset. Numerical features are embedded into two-dimensional coordinates for visualization, and model performance is quantified using scaled prediction errors. Among the evaluated architectures, CrabNet consistently demonstrates the highest predictive accuracy across all targeted electrochemical properties.

Compared to prior studies utilizing this dataset, and despite the absence of structural descriptors, the composition-based CrabNet model for gravimetric capacity outperforms previously reported structural models, such as Extreme Tree Regression. Furthermore, its performance remains competitive with Light Gradient Boosting Machine and Deep Neural Network models \cite{zhang2024interpretable} that incorporate structural data. While these methods are inherently simplified representations of complex materials, these results validate the utility of composition-based approaches as an effective first-stage high-throughput screening tool. Such methods are particularly valuable when exploring vast compositional spaces with high precision or when working with datasets that lack comprehensive structural information.

The robustness of these models has been characterized by analyzing error distributions across working-ion groups, performing bootstrap resampling, and implementing clustering via t-SNE based on matminer-derived features. Additionally, specialized cross-validation strategies are provided to highlight the current limitations and generalization capabilities of the models. Further, the final trained model has been tested on a smaller dataset of unseen experimentally validated materials. The final model consistently ranks both theoretical and experimentally validated materials, demonstrating the practical value of ML-driven approaches to accelerate the discovery of electrode materials for next-generation energy storage, with final model weights made available to facilitate reproducibility and allow for comparative research.

\clearpage

\appendix
\section{Appendix}

\setcounter{figure}{0}
\setcounter{table}{0}

\counterwithin{figure}{section}
\counterwithin{table}{section}

\begin{figure}[h!]
    \centering
    \begin{subfigure}[t]{0.32\textwidth}
        \begin{overpic}[width=\textwidth]{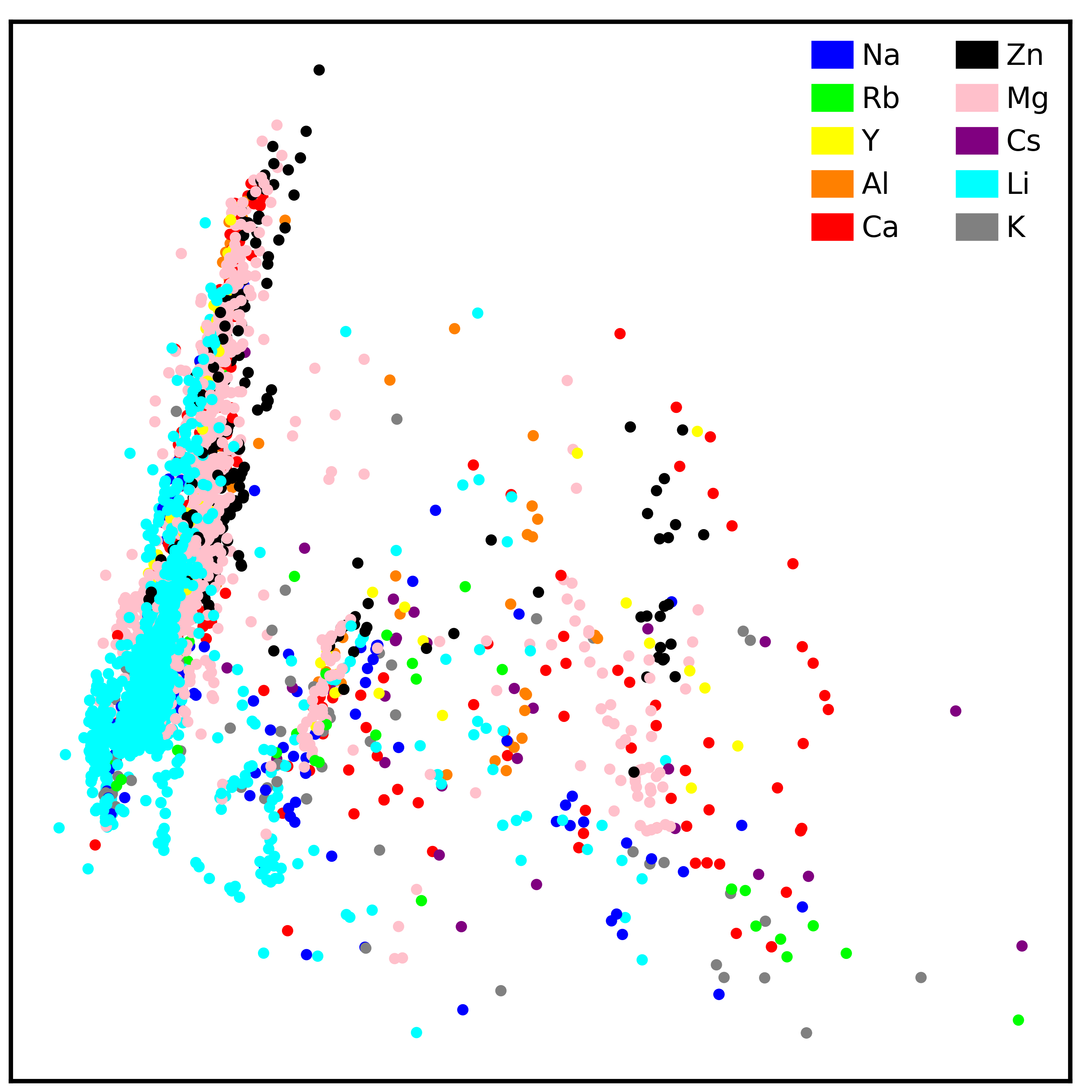}
            \put(0,100){\scriptsize\textbf{(a)}} 
        \end{overpic}
        \label{pca_modnet_ion}
    \end{subfigure}
    \hspace{-0.3em}
    \begin{subfigure}[t]{0.32\textwidth}
        \begin{overpic}[width=\textwidth]{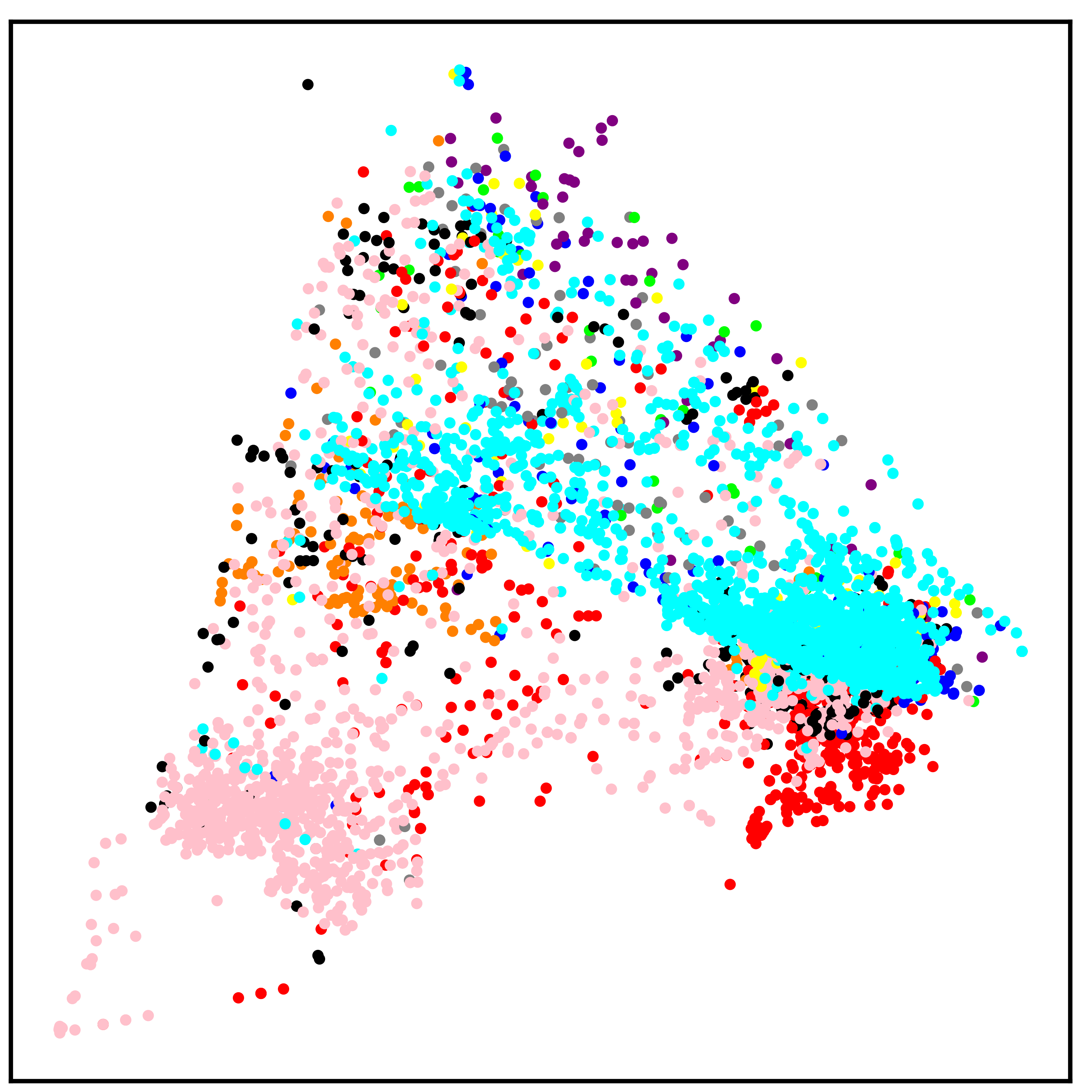}
            \put(0,100){\scriptsize\textbf{(b)}}
        \end{overpic}
        \label{pca_mat2vec_ion}
    \end{subfigure}
    \hspace{-0.3em}
    \begin{subfigure}[t]{0.32\textwidth}
        \begin{overpic}[width=\textwidth]{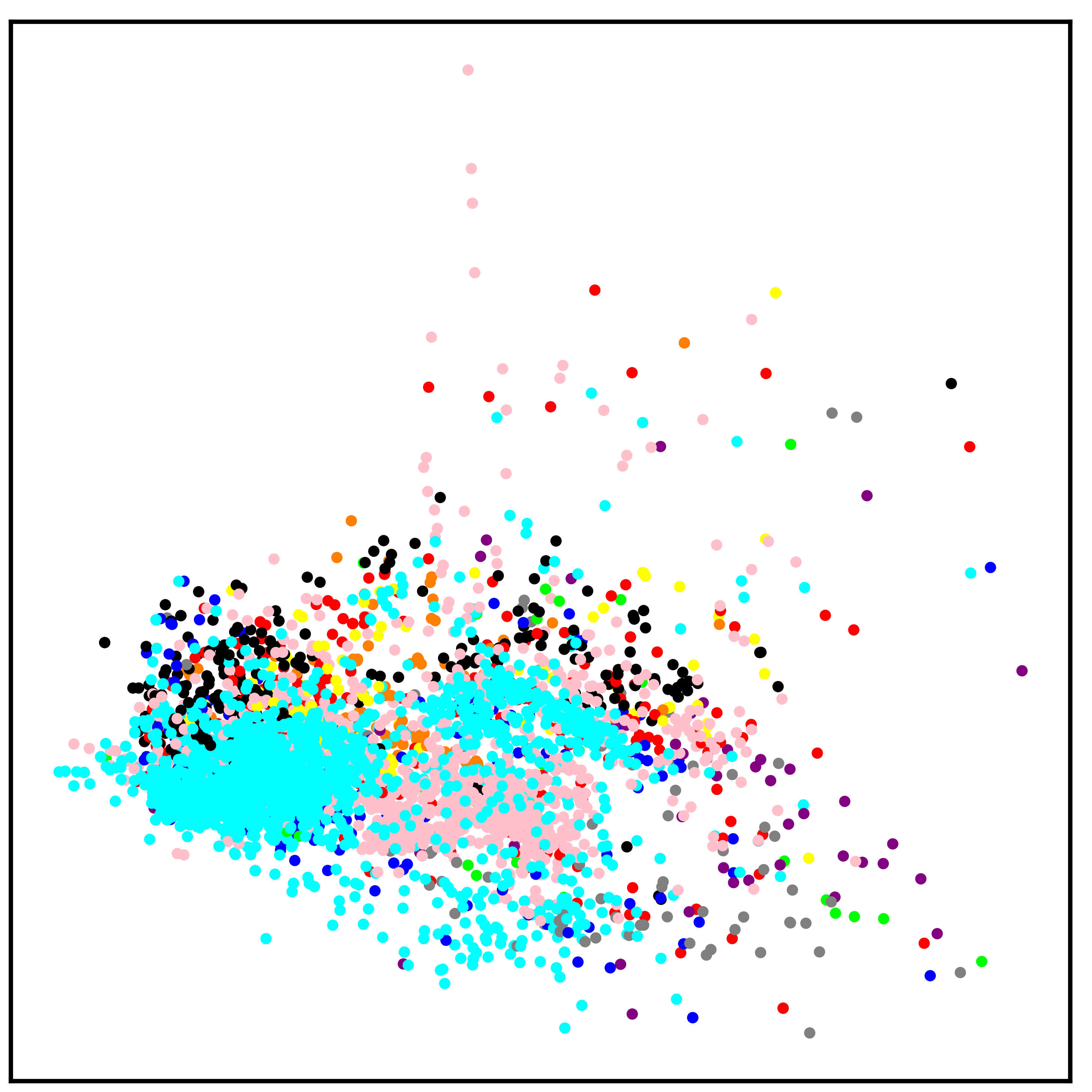}
            \put(0,100){\scriptsize\textbf{(c)}}
        \end{overpic}
        \label{pca_magpie_ion}
    \end{subfigure}

    \vspace{-0.3em}

    \begin{subfigure}[t]{0.32\textwidth}
        \begin{overpic}[width=\textwidth]{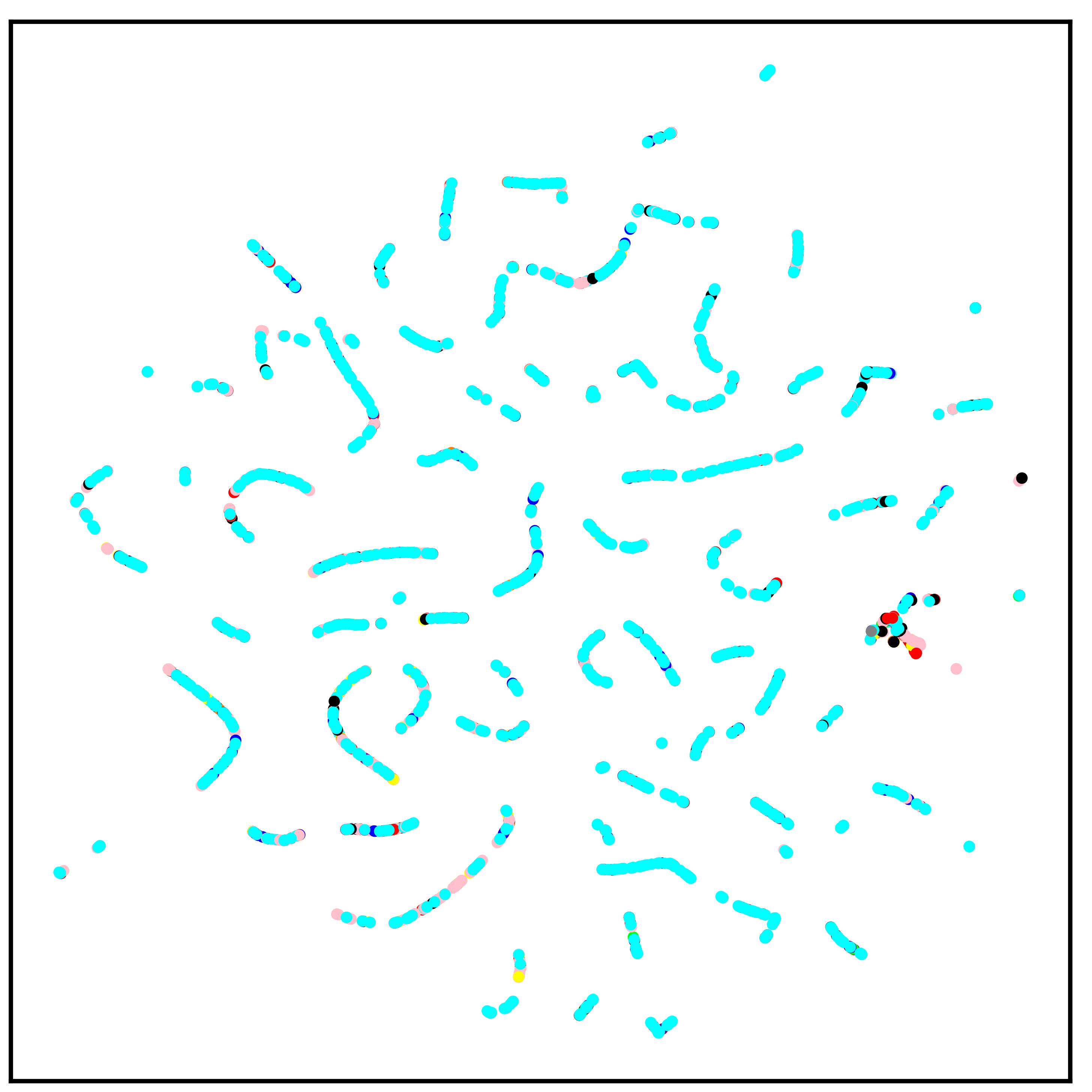}
            \put(0,100){\scriptsize\textbf{(d)}} 
        \end{overpic}
        \label{umap_modnet_ion}
    \end{subfigure}
    \hspace{-0.3em}
    \begin{subfigure}[t]{0.32\textwidth}
        \begin{overpic}[width=\textwidth]{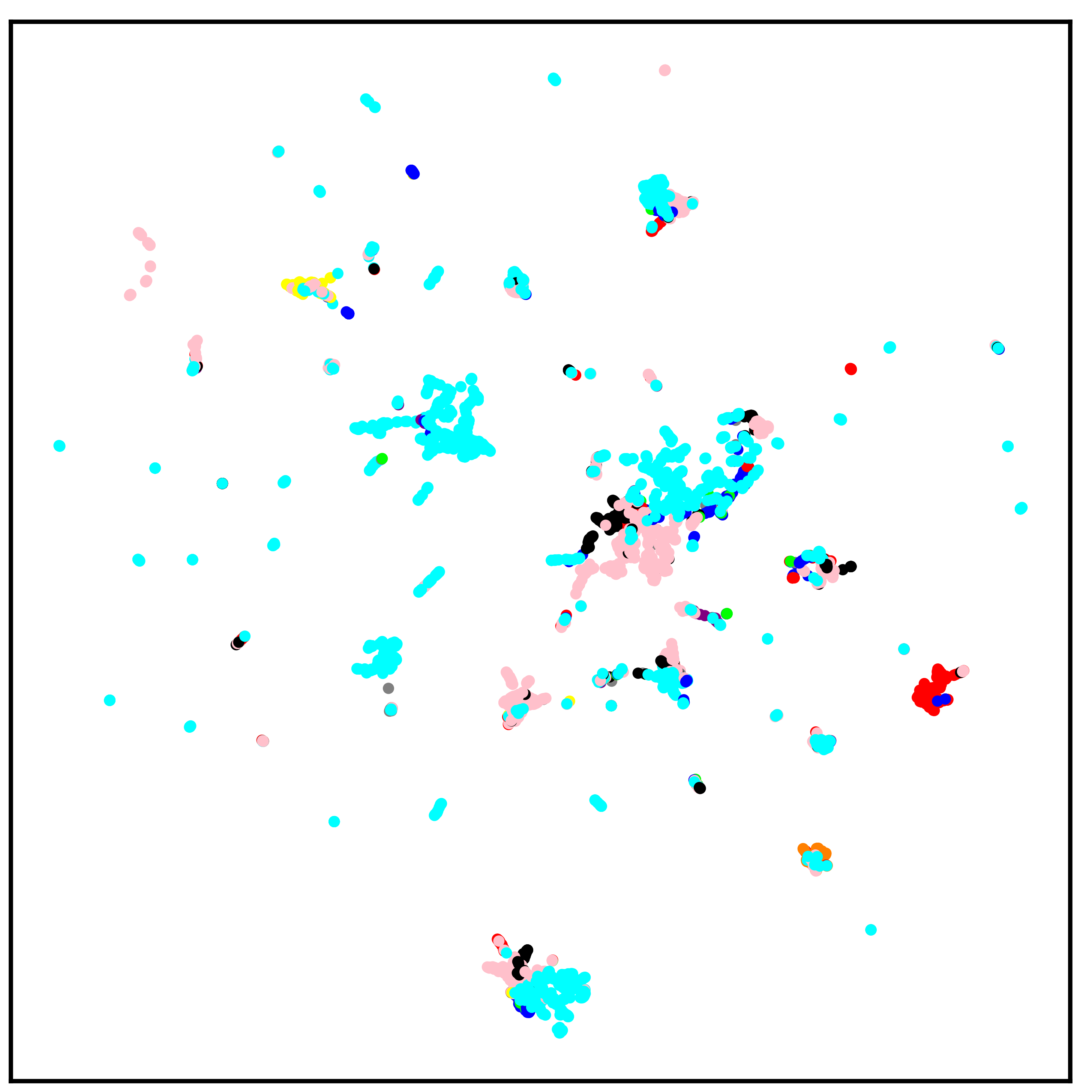}
            \put(0,100){\scriptsize\textbf{(e)}}
        \end{overpic}
        \label{umap_mat2vec_ion}
    \end{subfigure}
    \hspace{-0.3em}
    \begin{subfigure}[t]{0.32\textwidth}
        \begin{overpic}[width=\textwidth]{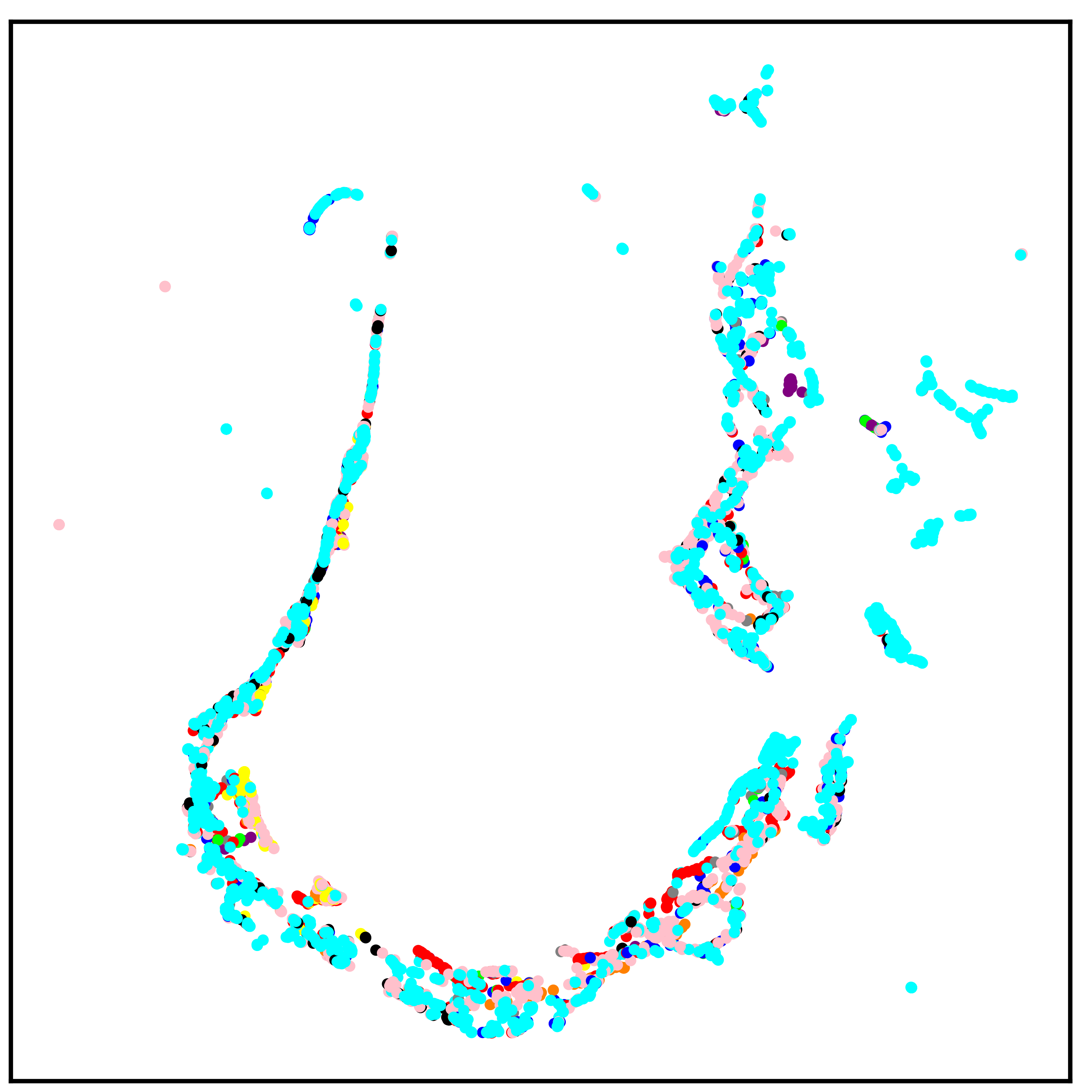}
            \put(0,100){\scriptsize\textbf{(f)}}
        \end{overpic}
        \label{umap_magpie_ion}
    \end{subfigure}
    \caption{5,574 candidate cathode materials from the materials project battery dataset are embedded using PCA (a, b, c) and UMAP (d, e, f) using input features from matminer (a, d), mat2vec (b, e), and magpie (c, f). Each of the plots of are colored by the working ion.}
    \label{over_ion_dropped}
\end{figure}

\begin{figure}[!h]
    \centering

    \begin{subfigure}[t]{0.6\textwidth}
        \begin{overpic}[width=\textwidth]{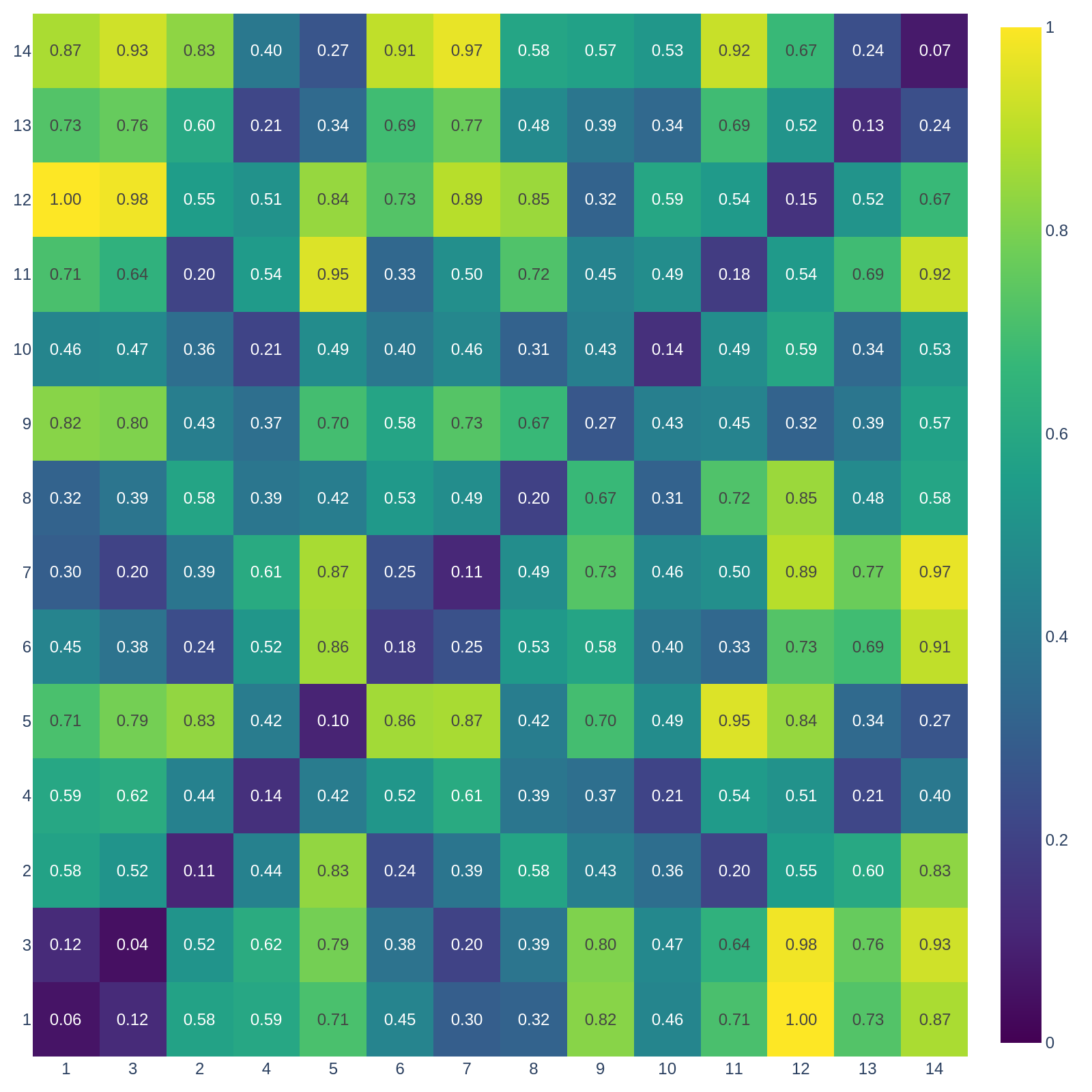}
            \put(0,100){\scriptsize\textbf{(a)}}
        \end{overpic}
    \end{subfigure}

    \vspace{0.6em} 

    \begin{subfigure}[t]{0.6\textwidth}
        \begin{overpic}[width=\textwidth]{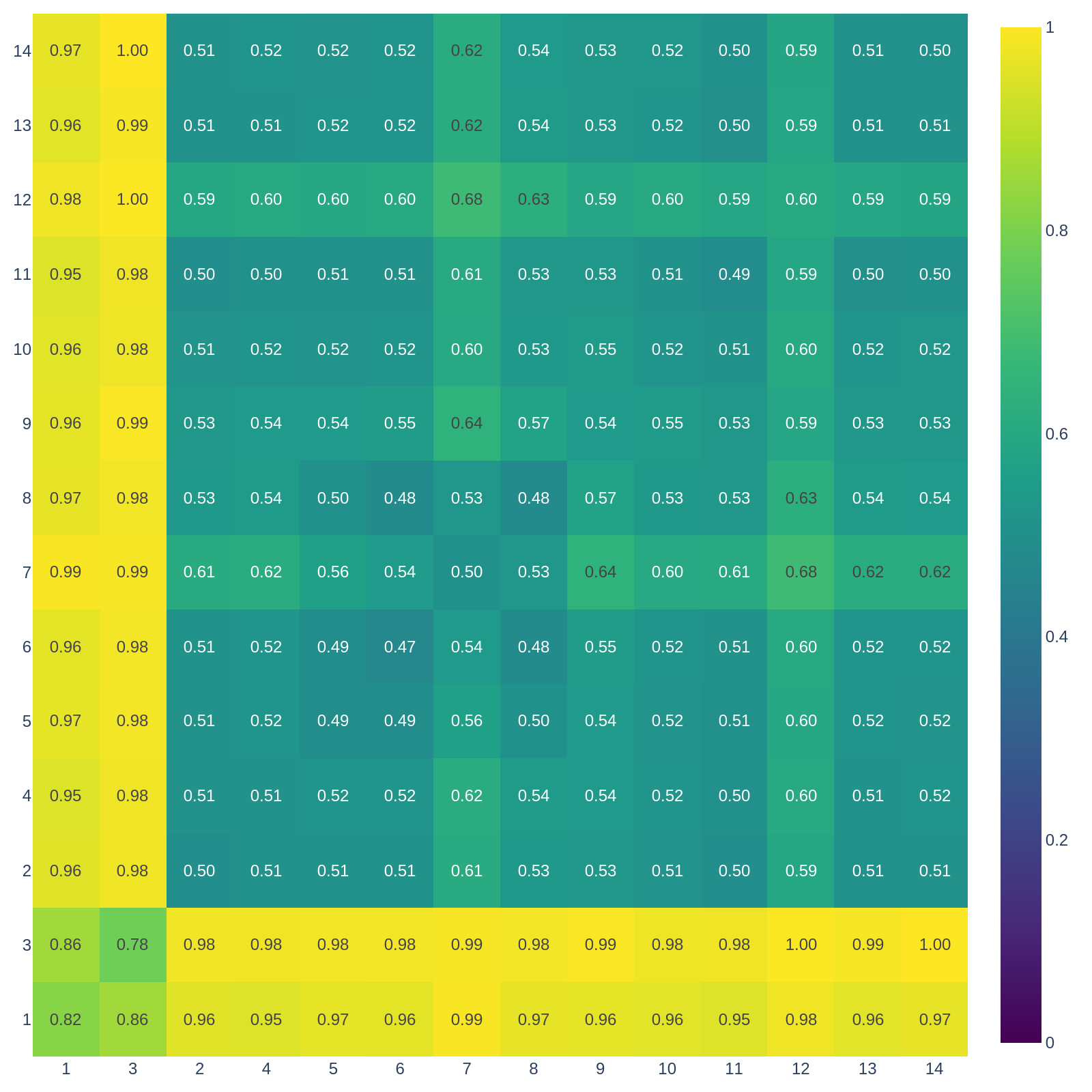}
            \put(0,100){\scriptsize\textbf{(b)}}
        \end{overpic}
    \end{subfigure}

    \caption{Average interpoint Euclidean distance matrix stratified across DBSCAN clusters. The distance matrices are from points in (a) 2D space from t-SNE and (b) original high-dimensional space. All distance values have been normalized across the respective distance matrix.}
    \label{dist_cluster}
\end{figure}

\begin{figure}[h!]
    \centering

    \begin{subfigure}[t]{0.6\textwidth}
        \begin{overpic}[width=\textwidth]{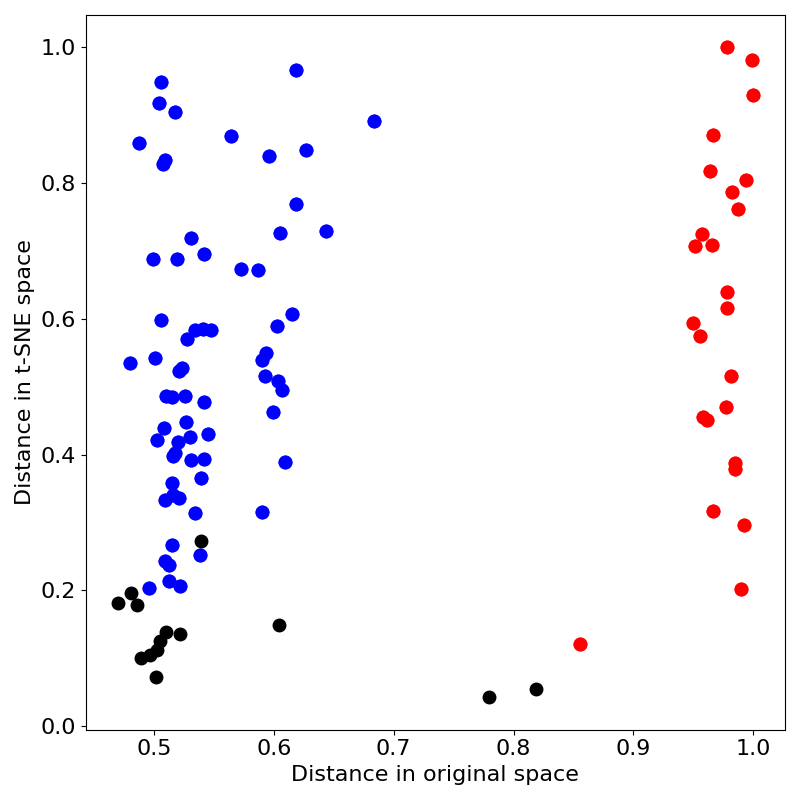}
            \put(0,100){\scriptsize\textbf{(a)}}
        \end{overpic}

    \end{subfigure}
    \vspace{0.3em}

    \begin{subfigure}[t]{0.6\textwidth}
        \begin{overpic}[width=\textwidth]{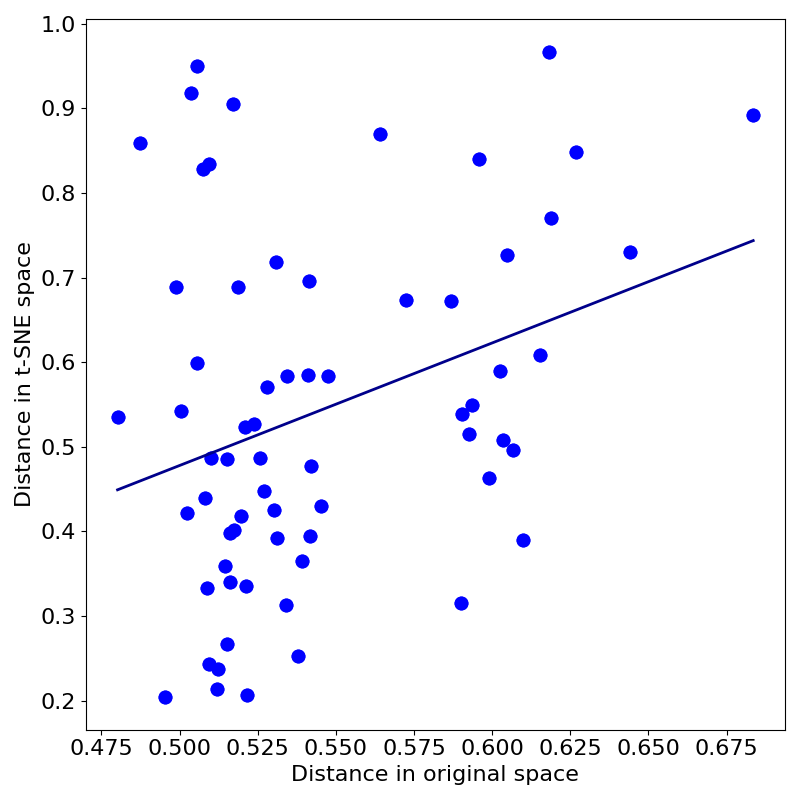}
            \put(0,100){\scriptsize\textbf{(b)}}
        \end{overpic}

    \end{subfigure}

    \caption{(a) Scatter plot demonstrating correlation between the two distance matrices in Figure~\ref{dist_cluster}, using Figure~\ref{dist_cluster}(b) and Figure~\ref{dist_cluster}(a) as the $x$ and $y$ intercepts respectively. Points associated with clusters 1 and 3 are shown in red, with the remaining inter-cluster distance values shown in blue, and intra-cluster distances shown in black.  
    (b) A subset of the $x<0.8$ inter-cluster distances with a linear regression best fit, yielding an $R^2$ value of 0.22.}
    \label{corr_dist}
\end{figure}

\newcolumntype{C}{>{\centering\arraybackslash}p{0.05\textwidth}}
\newcolumntype{P}{>{\centering\arraybackslash}p{0.08\textwidth}}

\begin{table}[t!]
\caption{Testing errors (SMAE) on the target properties for the different models (MODNet, CrabNet, RF@Magpie) from LOCO cross-validation over the clusters identified by DBSCAN. The best result for each column is highlighted in bold. The control model assigns the arithmetic mean of the training-set labels to all samples.}
\renewcommand{\arraystretch}{1.1}
\setlength{\tabcolsep}{4pt} 

\resizebox{\textwidth}{!}{%
\newcolumntype{C}{>{\centering\arraybackslash}p{0.05\textwidth}}
\newcolumntype{P}{>{\centering\arraybackslash}p{0.14\textwidth}}

\begin{tabular}{
c c
| C C C C C C C
 C C C C C C C
}
\hline\hline
\multirow{2}{*}{Property} & \multirow{2}{*}{Model}
& \multicolumn{14}{c}{Cluster} \\
\rule{0pt}{2.6ex}
& & 1 & 2 & 3 & 4 & 5 & 6 & 7 & 8 & 9 & 10 & 11 & 12 & 13 & 14 \\
\multirow{1}{*}{\makecell{Count}}
&  & 180 & 201 & 99 & 302 & 272 & 596 & 225 & 808 & 1025 & 466 & 418 & 516 & 277 & 189 \\
\hline
\multirow{4}{*}{\makecell{Gravimetric\\ Capacity\\(mAh/g)}}
& MODNet & \textbf{2.88} & 0.27 & \textbf{1.31} & 0.29 & 0.36 & 0.29 & 0.20 & 0.35 & 0.31 & \textbf{0.34} & 0.25 & 0.34 & 0.29 & 0.27 \\
& CrabNet & 3.44 & \textbf{0.26} & 1.37 & \textbf{0.24} & \textbf{0.29} & \textbf{0.25} & \textbf{0.19} & \textbf{0.31} & \textbf{0.22} & 0.30 & \textbf{0.20} & \textbf{0.22} & \textbf{0.27} & \textbf{0.26} \\
& RF@Magpie & 5.16 & 0.58 & 2.97 & 0.56 & 0.51 & 0.45 & 0.33 & 0.49 & 0.56 & 0.60 & 0.47 & 0.54 & 0.51 & 0.49 \\
& Control & 5.57 & 0.72 & 3.42 & 0.73 & 0.59 & 0.54 & 0.37 & 0.53 & 0.64 & 0.69 & 0.61 & 0.51 & 0.73 & 0.54 \\
\hline
\multirow{4}{*}{\makecell{Volumetric\\ capacity\\(mAh/cm$^3$)}}
& MODNet & \textbf{2.66} & 0.31 & 1.62 & 0.36 & 0.37 & 0.28 & 0.18 & 0.33 & 0.38 & 0.37 & 0.30 & 0.36 & 0.40 & 0.30 \\
& CrabNet & 2.75 & \textbf{0.27} & \textbf{1.41} & \textbf{0.29} & \textbf{0.32} & \textbf{0.24} & \textbf{0.17} & \textbf{0.27} & \textbf{0.24} & \textbf{0.35} & \textbf{0.26} & \textbf{0.24} & \textbf{0.33} & \textbf{0.30} \\
& RF@Magpie & 3.76 & 0.59 & 2.33 & 0.62 & 0.55 & 0.43 & 0.38 & 0.47 & 0.63 & 0.62 & 0.56 & 0.57 & 0.56 & 0.49 \\
& Control & 4.70 & 0.90 & 2.98 & 0.91 & 0.57 & 0.50 & 0.33 & 0.50 & 0.81 & 0.77 & 0.74 & 0.77 & 0.84 & 0.69 \\
\hline
\multirow{4}{*}{\makecell{Average\\ Voltage\\(V)}}
& MODNet & 1.14 & 0.28 & 1.40 & \textbf{0.50} & \textbf{0.49} & \textbf{0.32} & 1.45 & 0.57 & 0.48 & 0.58 & 0.40 & 0.82 & \textbf{0.29} & 0.27 \\
& CrabNet & \textbf{0.89} & \textbf{0.31} & \textbf{1.28} & 0.50 & 0.51 & \textbf{0.32} & \textbf{1.43} & \textbf{0.57} & \textbf{0.46} & \textbf{0.48} & \textbf{0.40} & \textbf{0.73} & 0.32 & \textbf{0.26} \\
& RF@Magpie & 1.47 & 0.49 & 1.66 & 0.65 & 0.92 & 0.58 & 1.95 & 0.77 & 0.67 & 0.81 & 0.52 & 0.96 & 0.52 & 0.55 \\
& Control & 0.99 & 0.77 & 2.45 & 0.91 & 0.84 & 0.72 & 1.80 & 0.90 & 1.90 & 0.83 & 0.83 & 1.19 & 0.80 & 0.73 \\
\hline\hline
\end{tabular}%
}
\label{tab:spcv1re}
\end{table}

\begin{table}[h!]
\caption{Testing errors of stratified 5-fold cross-validation from Figure \ref{tsne_dbscan_labels} and the average pairwise point distances in both original feature space (normalized values) and 2D space from t-SNE. The best result for each column is highlighted in bold. The control model assigns the arithmetic mean of the training-set labels to all samples.}

\newcolumntype{C}{>{\centering\arraybackslash}p{0.05\textwidth}}
\newcolumntype{P}{>{\centering\arraybackslash}p{0.05\textwidth}}

\begin{tabular}{
c c
| C C C C C c
}
\hline\hline
\multirow{1}{*}{Property} & \multirow{1}{*}{Model}
 & Fold 1 & Fold 2 & Fold 3 & Fold 4 & Fold 5 & Average \\
\hline
\multirow{4}{*}{\makecell{Gravimetric\\ Capacity\\(mAh/g)}}
& MODNet & 0.662 & 0.396 & 0.347 & 0.391 & 0.445 & 0.448 \\
& CrabNet & 0.541 & 0.339 & 0.305 & 0.291 & 0.465 & \textbf{0.388} \\
& RF@Magpie & 1.030 & 0.687 & 0.642 & 0.634 & 0.706 & 0.717 \\
& Control & 1.115 & 0.901 & 1.070 & 0.960 & 0.738 & 0.957 \\
\hline
\multirow{4}{*}{\makecell{Volumetric\\ Capacity\\(mAh/cm$^3$)}}
& MODNet & 0.674 & 0.445 & 0.369 & 0.355 & 0.525 & 0.466 \\
& CrabNet & 0.640 & 0.397 & 0.312 & 0.296 & 0.535 & \textbf{0.436} \\
& RF@Magpie & 1.156 & 0.829 & 0.616 & 0.565 & 0.649 & 0.723 \\
& Control & 1.130 & 0.978 & 0.901 & 0.807 & 0.835 & 0.930 \\
\hline
\multirow{4}{*}{\makecell{Average\\ Voltage\\(V)}}
& MODNet & 1.012 & 0.641 & 0.456 & 0.382 & 0.380 & 0.574 \\
& CrabNet & 0.995 & 0.329 & 0.421 & 0.374 & 0.391 & \textbf{0.562} \\
& RF@Magpie & 1.281 & 0.945 & 0.746 & 0.532 & 0.518 & 0.804 \\
& Control & 1.326 & 1.042 & 0.885 & 0.903 & 0.516 & 0.910 \\
\hline
\multicolumn{2}{c|}{Distance in original space} & 1.393 & 0.903 & 1.050 & 1.134 & 1.165 &  \\
\multicolumn{2}{c|}{Distance in 2D t-SNE space} & 14.484 & 12.755 & 27.484 & 18.721 & 13.640 &  \\
\hline\hline
\end{tabular}%

\label{tab:spcv2re}
\end{table}

\begin{table}[t!]
\centering
\caption{The selected compositions and their original sources chosen for the experimental holdout validation set, the Element Mover's Distance (ElMD) score to the most similar material in the training set, the experimentally reported and trained CrabNet model predictions are given for each of the three target labels, with the MAE of each target label given beneath.}

\setlength{\tabcolsep}{4pt} 
\resizebox{\textwidth}{!}{
\begin{tabular}{lccccccc}
\toprule
\hline
\multirow{2}{*}{\textbf{Material Formula}} & \multirow{2}{*}{\textbf{ElMD}} & \multicolumn{2}{c}{\textbf{Gravimetric Cap. (mAh/g)}} & \multicolumn{2}{c}{\textbf{Volumetric Cap. (mAh/cm$^3$)}} & \multicolumn{2}{c}{\textbf{Average Voltage (V)}} \\
\cmidrule(lr){3-4} \cmidrule(lr){5-6} \cmidrule(lr){7-8}
& & \textbf{Exp.} & \textbf{Pred.} & \textbf{Exp.} & \textbf{Pred.} & \textbf{Exp.} & \textbf{Pred.} \\
\midrule

$\text{LiNi}_{0.33}\text{Mn}_{0.33}\text{Co}_{0.33}\text{O}_2$ \cite{lee2004synthetic} & 0.15 & 160.0 & 145.2 & 955.2 & 703.7 & 3.75 & 2.45 \\

$\text{LiNi}_{0.8}\text{Co}_{0.1}\text{Mn}_{0.1}\text{O}_2$ \cite{marker2019evolution}    & 0.05 & 240.0 & 188.5 & 963.5 & 885.0 & 3.80 & 2.42 \\

$\text{Li}_{1.2}\text{Ni}_{0.2}\text{Mn}_{0.6}\text{O}_2$ \cite{hong2010structural}    & 0.65 & 238.0 & 215.3 & 1002.9 & 794.6 & 3.25 & 2.38 \\

$\text{Li}_{1.2}\text{Mn}_{0.5}\text{Ti}_{0.3}\text{O}_{1.9}\text{F}_{0.1}$ \cite{acs2026tailored}   & 1.24 & 253.0 & 214.4 & 997.9 & 733.5 & 3.25 & 2.61 \\

$\text{Na}_{0.44}\text{MnO}_2$ \cite{cao2011reversible}                               & 0.29 & 128.0 & 105.0 & 544.3 & 462.1 & 2.75 & 1.55 \\

$\text{Na}_{0.67}\text{Ni}_{0.33}\text{Mn}_{0.67}\text{O}_2$ \cite{pamidi2023single}
& 0.24 & 125.0 & 118.0 & 537.5 & 485.2 & 3.60 & 2.22 \\
    
$\text{Na}_3\text{V}_2(\text{PO}_4)_2\text{F}_3$ \cite{camacho2025lattice}             & 1.60 & 128.0 & 128.0 & 384.0 & 405.8 & 3.90 & 2.51 \\

\midrule
\textbf{Mean Absolute Error (MAE)} & \textbf{--} & \multicolumn{2}{c}{22.5} & \multicolumn{2}{c}{137.0} & \multicolumn{2}{c}{1.17} \\
\hline
\bottomrule
\end{tabular}%
}
\label{tab:expvali}
\end{table}

\clearpage
\section*{Acknowledgment}

G.-M.R. is Research Director of the Fonds de la Recherche Scientifique - FNRS. Computational resources have been provided by the Consortium des Equipements de Calcul Intensif (CÉCI), funded by the Fonds de la Recherche Scientifique de Belgique (F.R.S.-FNRS) under Grant No. 2.5020.11 and by the Walloon Region. The present research benefited from computational resources made available on the Tier-1 supercomputer of the Fédération Wallonie-Bruxelles, infrastructure funded by the Walloon Region under grant agreement no. 1117545. C. J. H. thanks the FRS-FNRS for their support as part of the Fish4Diet project.

\noindent The author acknowledged the consideration of the high-energy cathode for one of the battery outputs in the Horizon Europe (HEU) project named STELLAR (Grant Agreement No. 101202298), funded by the European Climate, Infrastructure and Environment Executive Agency (CINEA) as a part of the European Commission (EC). The views and opinions expressed are those of the author(s) and do not necessarily reflect those of the European Union or the granting authority.





%
\bibliography{science_template} 
\bibliographystyle{sciencemag}

%
%
%
%
%
%




\end{document}